\documentclass[journal,twocolumn]{IEEEtran}
\usepackage[english]{babel}
\usepackage[utf8]{inputenc}
\usepackage{amsmath}
\usepackage{cite,graphicx}
\usepackage{amsmath,amssymb,amsthm,bbm}
\usepackage{subcaption}

\usepackage{hyperref}
\hypersetup{
        colorlinks = true,
        citecolor=blue,
}

\usepackage[font=footnotesize]{caption} 
\usepackage{csquotes}
\usepackage{float}
\usepackage{comment}
\usepackage{algorithm}
\usepackage[noend]{algpseudocode}

\makeatletter
\def\BState{\State\hskip-\ALG@thistlm}
\makeatother

\makeatletter
\renewcommand*\env@matrix[1][*\c@MaxMatrixCols c]{%
  \hskip -\arraycolsep
  \let\@ifnextchar\new@ifnextchar
  \array{#1}}
\makeatother
\usepackage{csquotes}

\usepackage{titlesec}
\usepackage{mathrsfs}
\titlespacing*{\section}{0pt}{3mm}{1.5mm}
\titlespacing*{\subsection}{0pt}{2.5mm}{0pt}
\makeatletter
\newcommand*\dashline{\rotatebox[origin=c]{90}{$\dabar@\dabar@\dabar@$}}
\makeatother
\makeatletter
\newcommand*{\rom}[1]{\expandafter\@slowromancap\romannumeral #1@}
\makeatother

\newcommand{\norm}[1]{\left\Vert#1\right\Vert}

\newtheorem{mytheorem}{\bf Theorem} 
\newtheorem{myexample}{\it Example} 

\def\be{ \begin{equation} }
\def\ee{ \end{equation} }
\def\bea{ \begin{eqnarray} }
\def\eea{ \end{eqnarray} }

\def\b0{{\bf 0}}

\ifCLASSOPTIONonecolumn
\else
  
\fi

\begin{document}

\title{Understand-Before-Talk (UBT): A Semantic Communication Approach to 6G Networks}
 \author{Shiva Raj Pokhrel and Jinho Choi
  \thanks{The authors are with Deakin University, Geelong, VIC 3220, Australia(e-mail: \{shiva.pokhrel, jinho.choi\}@deakin.edu.au). A preliminary version of this work is submitted for presentation in IEEE Wireless Communications and Networking Conference (WCNC) 2023.
  }
\vspace{-4 mm}}

\maketitle

\begin{abstract}
 \textcolor{black}{In Shannon theory, semantic aspects of communication were identified but considered irrelevant to the technical communication problems. Semantic communication (SC) techniques have recently attracted renewed research interests in 
 $\rm 6^{th}$ generation (6G) wireless, because they have the capability to support an efficient interpretation of the significance and meaning intended by a sender (or accomplishment of the goal) when dealing with multi-modal data such as videos, images, audio, text messages, and so on, which would be the case for various applications such as intelligent transportation systems where each autonomous vehicle needs to deal with real-time videos and data from a number of sensors including radars. 
 To this end, most of the emerging SC works focus on specific data types and employ sophisticated machine learning models including deep learning and neural networks. However, they could be impractical for multi-modal data possibly within a real-time constraint, relative to the purpose of the communication. A notable difficulty of existing SC frameworks lies in handling the discrete constraints imposed on the pursued semantic coding and its interaction with the independent knowledge-base, which makes reliable \textbf{semantic extraction} extremely challenging.
 Therefore, we develop a new hashing-based semantic extraction approach to SC framework, where our learning objective is to generate one time signatures (hash codes) using supervised learning for low latency, security and efficient management of the SC dynamics.  We first evaluate the proposed semantic extraction framework over large image data sets, extend it with domain adaptive hashing  and then demonstrate the effectiveness ``semantics signature" in bulk transmission and multi-modal data.}
\end{abstract}

{\IEEEkeywords
Semantic communication; 6G; Hashing}
\ifCLASSOPTIONonecolumn
\baselineskip 26pt
\fi

\section{Introduction} \label{S:Intro}

Nikola Tesla once said: ``\textit{When wireless is perfectly applied the whole earth will be converted into a huge brain, which in fact it is, all things being particles of a real and rhythmic whole},"  as Tesla envisioned today's mobile phones and future sixth-generation  (6G) networks back in 1926.

 In Shannon theory, semantic aspects of communication were considered irrelevant to the technical communication problems~\cite{ShannonBellLab}. \textcolor{black}{From earlier~\cite{ShannonBellLab} and recent works~\cite{9398576, kountouris2021semantics, strinati20216g, pokhrel2022learning} in the literature, and motivated by the above Tesla's quote, we found that 6G design innovations can get valuable suggestions by cautiously observing how our brain's processes environmental inputs gathered by our senses.
Our brain learns, in substance, from prior acts (and also from our historically collected understandings and culture) and executes complicated reasoning by using sustainable energy in a short time. Such \textit{semantic aspects} (messages have meaning, and they refer to or are correlated according to some systems with certain physical or conceptual entities~\cite{ShannonBellLab}), of communication become significant for 6G~\cite{uysal2021semantic}.}

\begin{figure}[t]
\begin{center}
\includegraphics[scale=0.26]{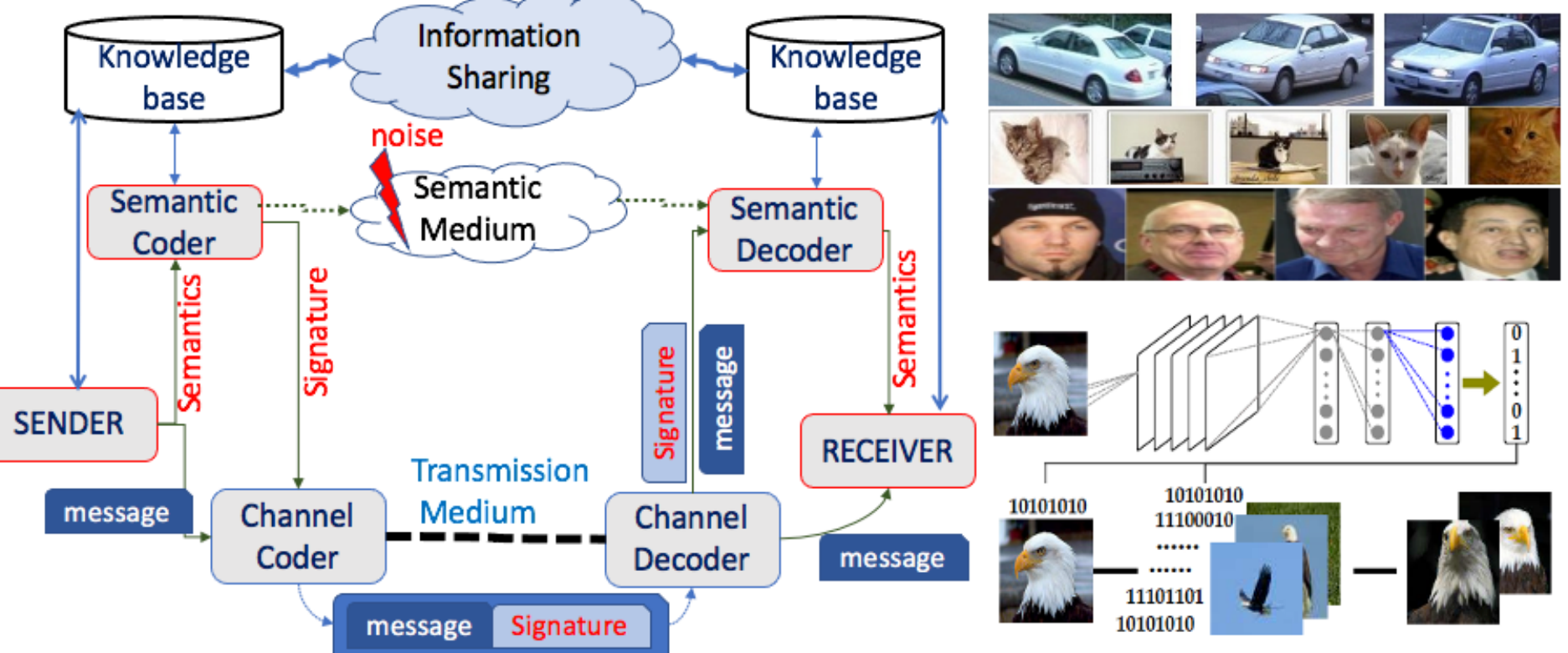}
\end{center}
\caption{\textcolor{black}{An abstract view of the proposed SC system. With semantic coding, we generate semantic signatures (by adopting hash codes). Images are fed and their corresponding hash are obtained by binarizing the activations with classification. For image recovery at receiver, the signatures are used as a query by the semantic decoder and every image in the knowledge base are compared based on the Hamming radius to generate semantic aspects of the images closest to the query.}}
\label{Fig:system}
\end{figure}

In 6G networks, inheriting these excellent human brain mechanisms  given by nature, should include \textit{semanticization and effectiveness in communication}s as the fundamental components for imminent innovations~\cite{yang2022semantic}.
As shown in Fig.~\ref{Fig:system}, such a mechanism allows us to distill the data that are strictly relevant for conveying (semantic) information from sender to receiver and focus on to clearly establish the purpose of communication. For example, data-aided sensing (DAS) proposed in~\cite{Choi_GPR} learns from prior acts and can be a valuable technique to considerably minimize the quantity of data to be transmitted (by decreasing irrelevant data transfer and using historically acquired experience).


Semantic communication, as illustrated in Fig.~\ref{Fig:system}, 
roots back to the three levels founded by \textbf{Shannon  and  Weaver}: 
\begin{itemize}
\item \textit{Technical (Level A)}: How correctly can the bits/packets be transmitted from the sender to the receiver?
\item   \textit{Semantics (Level B)}: How accurately do the bits/packets carry the required meaning from the sender to the receiver?
\item \textit{Effectiveness (Level C)}: How well does the meaning transmitted (conveyed) influence the intended course of action at the receiver?
\end{itemize}

At \textit{level A}, semantic aspects of communication are assumed to be irrelevant to the technical engineering problems. Shannon noted ``\textit{Frequently the messages have meaning; that is they refer to or are correlated according to some system with certain physical or conceptual entities. These semantic aspects of communication are irrelevant to the engineering problem}"~\cite{ShannonBellLab}. This is one of the fundamental assumptions of Shannon's theory. The relaxation of this fundamental assumption for communication between a sender and a receiver can jointly consider all three levels of problem, which can be referred to as \textit{semantic communication} (SC). In addition, \textcolor{black}{SC can borrow valuable insights from Sperber  and  Wilson’s \textit{inferential model} in communication~\cite{sperber1986relevance}-- `human cognition has a goal'}.

{\color{black}\subsection{Motivation} 

Conventional communication systems' actual transmission rates are getting close to the Shannon limit, and the remaining spectrum resources are being increasingly scarce~\cite{sana2022learning, yang2022semantic}.
The conversion of the transmission-before-understanding communication paradigm to the \textit{understanding-before-transmission} paradigm will be the key to SC's success in addressing bandwidth bottlenecks and spectrum scarcity. This is refer to as \textbf{Understand-Before-Talk (UBT)} in this paper. By doing this, semantic extraction can be included into the communication model to create SC for 6G~\cite{lan2021semantic}, which only permits the transmission of information that is relevant to the recipient and hides or reduces redundant data, relieving bandwidth pressure and improving privacy protection~\cite{yang2022semantic, zhang2022toward}. 
However, accuracy of semantic extraction has been highly challenging and non-trivial task in SC engineering.

One of the fundamental simplifying assumptions of Shannon's theory is that the information or the semanticized content is inherent in the transmitted message bits and that it is self-sufficient to retrieve the semantic information, once acquired by the receiver. Beyond the scope of Shannon's theory, a tightly coupled synchronization between sender and receiver is essential for successfully retrieving semantics at both ends. In contrast to the recent developments~\cite{yang2022semantic,  weng2021semantic, 8476247, 9252948, 9398576, 9450827, tucker2021emergent}, detailed later in Sec.~\ref{sec:IIB} and~\ref{sec:literature}, our objective is to develop an accurate semantic extraction approach for SC to facilitate joint optimization (\textit{information gathering, information dissemination,
and decision-making}) over 6G networks. 

We design a hashing-based semantic extraction framework for generating signatures enabling a low-delay generic SC approach applicable across all  message types. Different from other works, we are motivated by the human communication's realities, which demonstrate that \textit{what we send materially is entirely insufficient in interpreting the actual allusions that the sender meant to mean by transmitting the message}. Main components essential for indicating or acting in semantic sense, are:
i) surroundings and situations, ii) knowledge sharing, iii) prior communications held (used in~\cite{Choi_GPR}), iv) using empathy by sender/receiver, etc.}

Our preliminary moves towards this end are the studies towards employing federated learning~\cite{pokhrel2020fede} of connected vehicles for \textit{knowledge sharing}~\cite{pokhrel2022learning}, probabilistic inference based extraction~\cite{choi2022unified} and the data-aided sensing~\cite{Choi_GPR} (by taking into account the \textit{prior communications held}). Indeed, it is hard to ascertain that the information is contained just in the message sent, rather than which, we should understand messages to be a key to information access (a more sophisticated process) and this could be the core of 6G communication~\cite{nafria2009semantic, pokhrel2022learning}. 
\begin{table*}[t]
\caption{Existing Approaches and Challenges of Semantic Extraction for SC over 6G.\label{tab:SE}}
\centering
\begin{tabular}{c| c |c | c }
\hline
\textbf{Semantic extraction methods } & \textbf{Resolved issues} & \textbf{Impeding Challenges} & \textbf{Refs.}\\
\hline
\hline
Deep Learning based & Improved robustness over erroneous links;&  Partially studied semantic coding & \cite{xie2021deep,lee2019deep, xu2021wireless, hu2022robust, farsad2018deep, jiang2022deep} \\&Reduced payload length &But no semantic understanding; &\\
\hline
Reinforcement Learning based & Developed generalized framework;& Poor flexibility in the semantic extraction;& \cite{carpi2019reinforcement,  lu2022rethinking, kang2022task, lu2021reinforcement}\\
& Introduced new semantic metrics  &No multimodal and variable goal orientation&  \\
\hline
Knowledge Base driven & Goal-oriented semantic extraction & Lack adaptiveness for evolving communication goals & \cite{yang2021semantic, zheng2021knowledge, basu2014preserving} \\
\hline
Semantic Native and  & Developed contextual reasoning capabilities  & Largely theoretical with implementation issues & \cite{seo2021semantics, goodman2016pragmatic, choi2022unified, lazaridou2020emergent} \\
 Probabilistic Logic&Exploit \textit{entropy (clause)} in knowledge base & &  \\
\hline
Hashing-based approach & Proposed flexible, multimodal & Implementation focused partially studied details&  \\
 &  secure and variable goal solution& & This work \\
\hline
\end{tabular}
\end{table*}
{\color{black}\subsection{Impeding Challenges in SC}
\label{sec:IIB}
Despite the fact that SC has been viewed by many new approaches as a robust 6G technique, there are still several issues to be addressed.
We provide a synopsis of three typical challenges in SC and the relevant technologies and methodologies that are available for various uses. 

a) \textit{Accuracy of Semantic Extraction.} \noindent Semantic extraction~\cite{rachana2021literature} has been quite successful in different contexts, such as computer vision (e.g., semantic segmentation), semantic computing and semantic web~\cite{hitzler2021review} (similar to knowledge graph) for recommendation systems to facilitate intelligent and high quality user experience~\cite{patel2021present}. SC presents challenges for \textit{accurate semantic extraction}, particularly for the communication entities whose context may evolve individually, because all entities need to be strongly aligned semantic interpretation and representation.
Additionally, data communication in 6G has a high accuracy and a strong time-sensitive requirements (so called extreme URLLC~\cite{park2020extreme}).

In this paper, we focus on the challenges specific to the semantic extraction in SC. In Table~\ref{tab:SE}, the primary semantic extraction approaches are discussed along with the associated problems that have been solved, partially studied and impeding progress. Some researchers~\cite{xie2021deep,lee2019deep, xu2021wireless, hu2022robust, farsad2018deep, jiang2022deep} introduced deep learning enabled extraction methods known to be successful in the field of computer vision, natural language processing and speech processing. They have improved robustness in the low SNR regime. However, due to the lack of differentiable loss function, deep learning methods have inherent limitations. In other words, all of these works~\cite{xie2021deep,lee2019deep, xu2021wireless, hu2022robust, farsad2018deep, jiang2022deep} train neural networks using cross entropy and mean square error, which keeps the present efforts far from the intended practicability over SC~\cite{lu2022rethinking}. The semantic extraction are handled as a black box in such end-to-end designs, where the semantic channel encoders and decoders must be concurrently trained. Indeed, the efficiency of deep learning enabled extraction is difficult to gauge due to the lack of interpretability. 

In different contexts, reinforcement learning has been seen as a viable approach for dealing with the challenges associated with user-defined, task-specific, and perhaps even non-differentiable metrics~\cite{carpi2019reinforcement,  lu2022rethinking, kang2022task, lu2021reinforcement}. As summarized in Table~\ref{tab:SE}, learning the appropriate policy through interactions with the surroundings overcomes the limits of semantic metrics' differentiability~\cite{lu2021reinforcement}. However, it also increases training complexity~\cite{lu2022rethinking}. Training such a large model from start for a high-dimensional job becomes a significantly challenging task~\cite{yang2022semantic}. Furthermore, semantic information will be distinct for different tasks and therefore deep and reinforcement learning based methods are not applicable for multitask scenarios (may introduce irrelevant information and needs to be retrained, compromising system performance and resources)~\cite{yang2021semantic, zheng2021knowledge, basu2014preserving}.

Knowledge-based approaches~\cite{ zheng2021knowledge, yang2021semantic} resolve multitasking challenge but it is hard to generalize (due to the lack of explainable AI required for extracting semantic information hidden in deep nets). Furthermore, all of them (deep learning based, reinforcement learning based and Knowledge base driven)~\cite{xie2021deep,lee2019deep, xu2021wireless, hu2022robust, farsad2018deep, jiang2022deep, carpi2019reinforcement,  lu2022rethinking, kang2022task, lu2021reinforcement, yang2021semantic, zheng2021knowledge, basu2014preserving} are impractical for scenarios when semantics are time varying~\cite{seo2021semantics, ogden1923meaning, lazaridou2020emergent}. 

The semantic-native~\cite{seo2021semantics} and probabilistic logic~\cite{choi2022unified} SC system (with relevant insights from~\cite{seo2021semantics, goodman2016pragmatic, choi2022unified, lazaridou2020emergent}), indeed, has a high level of flexibility and adaptability, which is more in line with the evolving conceptualization of 6G. However, the studies~\cite{seo2021semantics, choi2022unified, lazaridou2020emergent}, once again, are preliminary and theoretical, and putting it into networks remains a tremendous task. With relevant insights from these works (c.f. Table~\ref{tab:SE}) understanding their implementation challenges and witnessing the practical success of hash-based signatures such as XMSS~\cite{hulsing2018xmss} in the postquantum cryptography, we take a major departure to exploit the capabilities \textit{one time signature} (OTS)~\cite{butin2017hash} by developing a hash-based semantic extraction for SC. 

b) \textit{Effectiveness of Metrics.} Research on semantic measurements is relatively poor and most SC assessments are primarily concerned with \textit{semantic error, age of information, and its value}. However, it is currently unclear how to estimate the quantity of semantic information and the required network resources. See~\cite{qin2021semantic,yang2022semantic} for further details. \noindent 

c) \textit{Quantification of Semantic Noise.} Semantic noise is a resultant of the mismatch between the sender and receiver's knowledge libraries. Most SC studies assume that the \textit{sender and receiver share the same invariant knowledge set}~\cite{xie2021deep} and the \textit{libraries of both parties remain consistent}~\cite{yang2021semantic}. However, each transceiver's background knowledge might grow independently, and data can even be intentionally attacked during download and transmission, leading to privacy and security issues. Therefore, relaxation of this fundamental assumption seems mandatory for further research~\cite{guler2018semantic}.} \noindent

\subsection{Semantic Extraction: Our Idea and Contributions}
{\color{black}{Hash functions transform arbitrary-length messages to fixed-length signatures. We mandate cryptographic hash-functions to be one-way (difficult to locate an ingredient in the signature of the input) and collision resistant (hard to discover two inputs that map to the same output). Hashing is one of the most successfully used cryptographic tools, with range of applications. While hashing is employed to handle arbitrary length messages in all practical signature schemes, it has been known (from the beginning of public key cryptography) that they can be used as sole building blocks for secure and meaningful communication. NIST has released SP 800-208~\cite{cooper2020recommendation} and standardized hash-based signature methods: LMS~\cite{mcgrew2019leighton} and SPHINCS+~\cite{bernstein2019sphincs+}. }} 

 Supervised hashing has been known to be successful for document and multimedia retrieval and representation learning~\cite{xia2014supervised}. Therefore, we develop a lightweight supervised hashing framework to enable flexible semantic extraction in SC of heterogeneous messages (texts, audio, images, video etc.) by generating a \textbf{semantics signature}.

Recent deep learning-enabled SC studies rely on shared background knowledge between sender and receiver, which is not practicable. We are inspired by the idea proposed in~\cite{tucker2021emergent} to communicate via discrete tokens derived from a continuous learned space. However, a considerable difficulty of SC lies in handling the discrete constraints imposed on the pursued semantic coding.

\textcolor{black}{As illustrated in Fig.~\ref{Fig:system}, we develop an UBT approach, which consists of i) \textit{understand-phase}, transforming  `\textit{semantics understanding}' of the message into a signature before  transmission by using knowledge base rules to convey the semantic information (sender wants to share with the receiver); and ii) \textit{talk-phase}, where the generated signature is either transmitted solely or piggybacked with the message, which will be used for reconstructing semantics and adapting knowledge base at the receiver (based on the shared knowledge base, ontology and rules). Unlike the existing  works, our learning mechanism is tightly coupled with hashing to generate optimal  signatures for transferring semantics from sender to the receiver.}

\textcolor{black}{Our main contributions in this paper are outlined as follows:
\begin{itemize}
    \item We design a simple semantic communication framework based on supervised hashing using signatures. Our design of semantic extraction and domain adaptation exemplifies the joint optimization of (\textit{information gathering, dissemination,
and decision-making}) over 6G networks.
    \item  We transform the underlying classification context into an optimization problem and then relax the problem using regularization for generating signatures. 
\end{itemize}
New insights developed in this work can be further extended to extensively investigate the benefits for performance, security and privacy of 6G networks.}

\section{Preliminaries and State-of-the-Art}
 Shannon's Information Theory (SIT) states that the transcription between message and symbol usually comprises  source encoding that reduces the redundancy contained by the message, followed by  channel encoding (structured redundancy to improve reliability). The concepts of SIT govern the rules and properties of source and channel encoding. 
This combination of source and channel encoding is called syntactic encoding~\cite{strinati20216g}, as it changes the form of the message  (but not the semantic content of the message). The sequence of symbols then is mapped into a physical electromagnetic wave or an acoustic wave, such that it is suitable for a physical transmission medium.

Traditional source messages in the SIT sense can be applied to encode semantic information, and there is a relationship between syntatic symbols and semantic messages~\cite{bao2011towards}. To elaborate, we consider semantic encoding at a source, which is randomly generating messages with probability $P_{M}(m), m\in M$, where $M$ represents the set of messages. Message $m$ is produced based on the ontology (and the rules from the knowledge base, see Fig~\ref{Fig:system}), which is converted into a sequence $x\in X$ of symbols, such that, $x=f(m)$. \textcolor{black}{The (syntatic) message entropy of the source can then be estimated (by applying Shannon's theory) as follows:
\begin{equation}
    H(M)=-\sum_{m\in M} P_{M}(m) \log_2 P_{M}(m),
\end{equation}
while the logical probability that the source generates the syntatic symbol $x$ is given by
\begin{equation}
  P(x)=\sum_{x:=f(m), m\in M} P_{M}(m).
\end{equation}
\begin{myexample}
Let $X = \{1,2,3\} = \{\mbox{car}, \mbox{cat},\mbox{man}\}$.
Then, $M$ is given by a set of images (top right panel of Fig.~\ref{Fig:system}), $m$ is an image that can be represented by a vector.
\end{myexample} 
Therefore, the semantic entropy associated with the symbols generated by the source can be computed as~\cite{strinati20216g}
\begin{equation}
    H(X)=-\sum_{x\in X} P(x) \log_2 P(x).
    \label{eqn:entr}
\end{equation}
Since $H(X)$ and $H(M)$ are different, considering the conditional entropies of $X$ to $M$ and $M$ to $X$, we can efficiently compute the mutual information between $X$ and $M$ as
\begin{align}
   B(M;X)&:=H(M)- H(M|X)\nonumber\\&=H(X)- H(X|M),
\end{align}
where $H(X|Y)$ is the conditional entropy of $X$ for given $Y$.
As shown in Fig.~\ref{Fig:system}, the semantic coder maintains the meanings linked to the produced syntatic messages. It was partially studied in~\cite{Basu2012} and reported that an encoder requires on average $B(M; X)$ bits only, as elaborated in (4), to encode a semantic aspects of a syntatic message generated by the source.
For example, the desired block coder for a semantic encoder at the source, as discussed in~\cite{Basu2012}, can be attained simply by employing $B$ bits semantics per syntatic message without any semantic error, which is referred to as \textit{semantic capacity}. For simplicity, we omit $(M; X)$ and use only $B$. It is worth noting that these $B$ bits enable us to piggyback semantic signature with the syntactic message packets during transmission.}

\subsection{Literature}
\label{sec:literature}
\begin{table}[t]
\caption{Comparing  relevant research works on semantic extraction.\label{tab:comparison_qualitative}}
\centering
\begin{tabular}{|c|| c |c | c | c| c|}
\hline
\textbf{Refs} & \textbf{Semantics} & \textbf{Channel} & \textbf{Approach}\\
\hline
\hline
\cite{9398576} & Text (Sentences) & AWGN & Deep Learning \\
\hline
\cite{8476247} & Text (Words) & binary symmetric& Bayesian game\\
&  &AWGN&  \\
\hline
\cite{weng2021semantic} & Speech Signal & AWGN, Rayleigh & Deep Learning \\
\hline
\cite{9252948} & Text  & Rayleigh, Rician  & Deep Learning \\
\hline
\cite{9450827} & Speech Signal & AWGN, Rayleigh&  \\
 &  &  Rician& Deep Learning \\
\hline
\cite{tucker2021emergent} & Continuous space & Noisy agent & Neural Agent \\
\hline
\textbf{This} & Text, Audio & AWGN & Discrete Hashing \\
\textbf{work} & Images, Video & Rayleigh, Rician & Supervised Learning \\
\hline
\end{tabular}
\end{table}

Of particular importance to this work are the initial research works~\cite{weng2021semantic, 8476247, 9252948, 9398576, 9450827, tucker2021emergent} on SC
systems (see the comparison in Tab.~\ref{tab:comparison_qualitative}).  Guler et al.~\cite{8476247} initialize research on SC
for text information to mitigate semantic errors.{\color{black} Weng et al.~\cite{weng2021semantic} developed a deep learning-based system to cope with various channel environments. Xie and Qin~\cite{9252948} applied deep learning to develop distributed SC system for text
transmission. Weng and Qin~\cite{9450827} proposed a mechanism to recreate the source
message by constructing the speech signal. The authors in~\cite{tucker2021emergent} developed a neural agent
that enables SC via discrete tokens extracted from continuous learning space.

We first formulate an optimization problem of the learning-based hashing by considering the underlying classification problem essential to generate semantics signature (see Sec.~\ref{Sec:2} A).  After that, we exploit \textit{the regularization  concept} from  large-scale  optimization to solve the relaxed optimization with the discrete binary constraints (Sec.~\ref{Sec:2} B), which helps us solve the problem iteratively (Sec.~\ref{Sec:2} C, Algorithm~1).} 

\section{Hashing-based Semantic Communication}\label{Sec:2}


{\color{black}
In 6G, we expect to support 
SC of heterogeneous messages by generating a semantics signature through semantic encoding. 
For semantic encoding, the task of learning binary hash codes is essential for the proposed framework to find semantics signatures. This task can be carried out by formulating a classification problem. 
In this section, we first provide the significance of the proposed framework and then formulate a classification problem and derive an iterative approach to solve the problem. 
}
 \subsection{Proposed SC Framework}
{\color{black} We introduce a new concept, a major departure from the traditional approach to designing and evaluating communication networks, by incorporating information semantics. The information semantics in our design are defined not simply as the meaning of the messages, but rather as their significance or importance, within a real-time constraint, and feasibly attributable to the goal of the communication. Our proposed signatures are the keys to i) efficiently synchronize the communication at all three levels and ii) consistently trigger updates of the knowledge bases. The \textbf{signature} employs a set of short binary codes to encode texts, images, videos, and other multimodal data while maintaining their original similarity. By using the signatures, due to the remarkable efficiency of pairwise comparison using the Hamming radius, the task of the nearest neighbour search can be readily performed on a large-scale knowledge base. Furthermore, 
such learned compact signatures (binary codes generated using data-driven hashing methods) can effectively and highly efficiently index and organize massive data in the knowledge base.

Importantly, the signatures learned can be viewed as nonlinearly synthesized feature vectors of the original data, which can jointly optimize information collection, distribution, and decision-making policies for a large scale autonomous network systems. The label information can be used to classify these feature vectors. One can observe in the proposed framework, how our design nonlinearly converts the original data into a binary space and then categorize the context-aware similarity in this space. By piggybacking the signatures with the message, we are synchronizing the communication at all three levels and advocating that the joint optimization (\textit{information gathering, information dissemination,
and decision-making}) is essential for the SC. The signature at the receiver side serves as a key to i) connect semantic decoder with the knowledge base,  ii) update the knowledge base (if needed) and iii) adjust the hashing function at the decoder to generate the original semantics. Next, we discuss knowledge-base and semantic medium in detail.

\emph{Knowledge Base}:
Facts, relations, and possible ways of reasoning are examples of aspects of real-world knowledge that can be understood, recognized and learnt by all communication parties -- forming the core elements of the Knowledge Base.
It is worth noting that the users at the source and destination don't always have to share the same knowledge base. However, both communication parties (sender and receiver) must be aware of the knowledge elements involved in their communication. To capture the rich semantics of knowledge elements as well as their complicated interaction, the sender and receiver must constantly maintain and continuously update their knowledge of the area models. As mentioned earlier, the updates of the knowledge-base in the proposed framework can be mostly triggered by the piggybacked signature while reconstructing semantics at the receiver end.

\begin{figure}[h]
    \centering
    \includegraphics[scale=0.4]{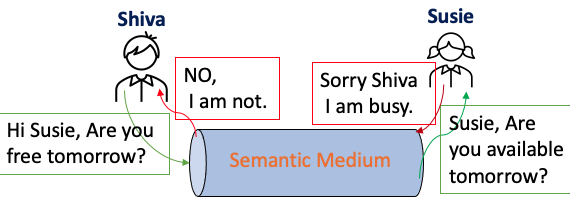}
    \caption{How semantic medium works?}
    \label{fig:semmedium}
\end{figure}

\emph{Semantic Medium}: There are primarily two kinds of transmission mediums in the proposed SC system. The first kind of medium is the physical wired/wireless medium, the so-called `transmission medium' where impairments on the transmitted signals are prevalent, due to fading, shadowing, distortion, and interference. Historically, most research efforts in communication have focused on alleviating the adverse effects of such impairments' over the physical transmission medium. There exists a second kind of medium in SC, referred to as `semantic medium', which is very different from the physical transmission medium. It doesn't focus on transmitting exactly the same message conveyed by the sender but rather aims to preserve the original meaning of the message.

An example of how semantic medium works is shown in Fig.~\ref{fig:semmedium}. Observe in Fig.~\ref{fig:semmedium}, although semantic medium does not accurately transmit Shiva's and Susie's original messages, it always preserves the sender's main intention. Therefore, if we evaluate the underlying communication success, there can be a technical engineering failure (at level A), but there is no semantic failure (at level B).  Semantic medium can be adversely impacted by the semantic noise generated due to misunderstanding, interpretation mistakes, or disruption in the estimation, retrieval or prediction of the information, which is one of the challenging problem in the SC. Perhaps the piggybacked signature can be used as a tool for ameliorating semantic noise in the medium as the label information can be used to classify their feature vectors.}\footnote{The observation of using signatures as a tool for detection and correction of semantic noise in the medium requires further investigation and is left for future work.}

\subsection{Learning Signatures by Fast Supervised Learning}

In recent years, real momentum has been building up for developing learning-based hashing techniques because they  support efficient processing and retrieval for higher-dimensional data
such as images, videos, documents, etc. \textcolor{black}{In the proposed SC framework, we often deal with high-dimensional data. Therefore we adopt classification information to learn binary hash codes (i.e., semantics signatures). We consider a discrete hashing mechanism consisting of $n$ message samples  $\mathbb X=\{\textbf{x}_i\}_1^n$,
where our objective is to learn a set of binary codes $\mathbb B=\{\textbf{b}_i\}_1^n$ to preserve their similarities in the semantics signature $\textbf{b}_i$ of  $B$ bits. More importantly, to leverage supervised learning, we inherit binary coding in linear classification with a hypothesis that optimal binary codes are beneficial for such classification problems. }

\textcolor{black}{We adopt vectors $\textbf{w}_k\in \mathbb R^{B\times 1}$ to perform the classification for class $k
\in \{1, 2, \dots c\}$, and formulate the underlying classification problem as}
follows:
\begin{equation}
    \min_{\mathbb B, \mathbb W, \mathbb H}\sum_{i=1}^n Loss(\textbf{b}_i, \textbf{y}_i\mathbb W^\intercal) +\alpha\norm{\mathbb W}^2,
    \label{eqn:main}
\end{equation}
\textcolor{black}{where 
$\mathbb Y=\{\textbf{y}_i\}_1^n$ is the ground truth matrix (if $\textbf{x}_i$ is of class $k$, $y_{ki}=1$ otherwise $0$), $Loss(\textbf{b}_i, \textbf{y}_i\mathbb W^\intercal)$ is the loss function, 
$\alpha$ is the regularization parameter, and 
${\mathbb W} = [ \textbf{w}_1 \ \cdots \  \textbf{w}_c]$} as{\color{black}
\begin{equation}\label{eqn6}
    \sum_{i=1}^n Loss(\textbf{b}_i, \textbf{y}_i\mathbb W^\intercal):=\norm{\mathbb B-\mathbb Y\mathbb W}^2 .
\end{equation}}
Here, $\mathbb H= {\rm sgn} (F(\textbf{x}))$ is the Hash function for encoding $\textbf{x}$ by  $B$ bits, where ${\rm sgn} (\cdot)$ outputs $1$ for all +ve numbers, otherwise $-1$. Observe in~\eqref{eqn:main} that we need 
\[\textbf{b}_i={\rm sgn}(F( {\bf x}_i)), i \in \{1,2,\dots,n\},\] even though it appears that we are only interested  in learning the hashing function $F(\textbf{x})$. Moreover, along the lines of~\cite{kulis2009learning}, removing $\textbf{b}_i$ for eliminating auxiliary variables and constraints is not possible here (as $\textbf{b}_i$ is of central focus). Also, such an approach leads to increasing computational complexity, which becomes not only very difficult but also more sluggish to minimize~\eqref{eqn:main}. 
\textcolor{black}{With the ideas from~\cite{shen2015supervised}, we developed a significantly more effective and efficient approach as solving~\eqref{eqn:main} is complicated (NP-hard) even with discrete $\textbf{b}_i$. A simple relaxation to this end would be obtained
by assuming that $\textbf{b}_i=F({\bf x}_i)$ is continuous, which is applied in most existing hashing schemes. However, the obtained solutions are sub-optimal, although this relaxation simplifies the complexity and makes~\eqref{eqn:main} easier to solve. Therefore, we exploit the different ideas of \textit{regularization from large-scale optimization} using a primal-dual approach to solve the problem more efficiently by keeping the discrete binary constraints of the semantcis signature $\textbf{b}_i$~\cite{7873258}.}

\subsection{Relaxation using Regularization}
Based on the above discussion and insights form~\cite{wang2014role}, by using penalty parameter, $\mathcal P$, \eqref{eqn:main} can be approximated 
as follows:
\begin{align}
    \label{eqn:main2}
 &\min_{\mathbb B, \mathbb W, \mathbb H}\Big( \mathcal P \sum_{i=1}^n \norm{\textbf{b}_i-F({\bf x}_i)}^2 \nonumber\\ 
 & \qquad + \sum_{i=1}^n Loss(\textbf{b}_i, \textbf{y}_i\mathbb W^\intercal)+\alpha\norm{\mathbb W}^2\Big)\\
& \mbox{subject to }\  \textbf{b}_i \in\{-1,1\}^B.\nonumber
\end{align}
We can see that~\eqref{eqn:main2} is non-convex and still highly difficult to minimize. Therefore, for tractability, we adopt a fixed point formulation where we keep two-variable fixed while computing one and thus solve~\eqref{eqn:main2} one by one in an iterative fashion. To this end, we develop an iterative procedure, and the design details are provided next (see Algorithm~1).

\begin{algorithm}[h]
\label{algorithm1}
\begin{algorithmic}[1]
\Procedure {Initialize}{}
\State \textbf{Input:} $\{ {\bf x}_i\}_{i=1}^n$ data sets (training); 
$B$-bit code length \State $m$ sample points, $\mathcal I$ Max. iterations; Parameters ($\alpha, \mathcal P$)
\EndProcedure
\Procedure{\textcolor{black}{Learning with Training}}{}
\State Random selection of $m$ samples
\State $\{ \hat {\bf x}_i\}_{i=1}^m$ from the training set  $\mathbb X=\{ {\bf x}_i\}_{i=1}^n$
\State Get mapped data $\phi(\textbf{x})$ from radial basis kernel
\State Random initialization vector: $\forall_i, \textbf{b}_i\in \{-1, 1\}^{B}$
\While{($\mathcal I$ iterations \textbf{or} Convergence)}
\State Fixed $\mathbb{B}$ and compute $\mathbb{W}$ using multiclass SVM
\State Projection matrix $\mathbb{Q}={(\phi(\mathbb{X})\phi(\mathbb{X})^{\intercal}})^{-1}\phi(\mathbb{X}){\mathbb{B}^{\intercal}}$
\State Learn $\textbf{b}_i$ iteratively
\State {$\textbf{b}_i={\rm sgn}\big(F( {\bf x}_i)+\frac{1}{2\mathcal P}\sum_{k=1}^c\textbf{w}^{(ki)}\big)$}
\EndWhile
\EndProcedure
\Procedure{Output}{}
\State Signatures $\mathbb B=\{\textbf{b}_i\}_1^{B\times n}$; Hash function $\mathbb H$
\EndProcedure
   
\caption{Signature Learning by Discrete Hashing}
\end{algorithmic}

\end{algorithm}

\section{Signature Learning Algorithm and Analysis}

The important design details of Algorithm~1 is explained as follows. \textcolor{black}{ By using the fixed-point formulation, for given $\mathbb{B}$ in~\eqref{eqn:main2},  $\mathbb{Q}$ is the projection matrix that can be computed by regression and is  (Algorithm~1, step 11) given by
\begin{equation}
    \mathbb{Q}={(\phi(\mathbb{X})\phi(\mathbb{X})^{\intercal}})^{-1}\phi(\mathbb{X}){\mathbb{B}^{\intercal}}.
    \label{eq:proj}
\end{equation} We find that any suitable linear or nonlinear learning algorithm can be used for $F(\textbf{x})$.} For our evaluation, we use a simple nonlinear $F(\textbf{x})=\mathbb{Q}^{\intercal}\phi(\textbf{x})$, where $\phi(\textbf{x})$ is a vector realized using radial basis kernel function of width $d$:
\begin{eqnarray}
\phi(\textbf{x})=\Big[\exp\left(-\frac{\norm{\textbf{x}-\hat {\bf x}_1}^2}{d}\right) ,\dots, \exp\left(-\frac{\norm{\textbf{x}-\hat {\bf x}_m}^2}{d}\right)\Big]^{\intercal}.\nonumber
\end{eqnarray}
\textcolor{black}{Here, $\{ \hat {\bf x}_i\}_{i=1}^m$ is obtained from a random selection of $m$ samples from the training set  $\{ {\bf x}_i\}_{i=1}^n$ (see Algorithm~1, Steps 5-6). In Algorithm~1, step 10, for a given $\mathbb{B}$, we compute the classification $\mathbb{W}$ by using multiclass kernel-based support vector machine (SVM) formulation of~\eqref{eqn:main2}.\footnote{We use LIBLINEAR -- a library for large linear classification, available at https://web.stanford.edu/~kimth/cs229/ps2/liblinear-1.8. The hinge loss is considered in loss function for training classifiers, in particular for the SVM.} In steps 12-13 of Algorithm 1, we reconsider a fixed point formulation accounting all variables with hinge loss, but $\textbf{b}_i$ is fixed. After relevant simplification, we observe that~\eqref{eqn:main2} can then be expressed as
\begin{align}
    \label{eqn:main222}
 &\min_{{\bf b}_i} \norm{\textbf{b}_i-F({\bf x}_i)}^2 - \frac{1}{\mathcal P}\sum_{i=1}^c (\textbf{w}^{(ki)\intercal}{y}^{(ki)}+{\bf b}_i)\\
& \mbox{subject to }\  \textbf{b}_i \in\{-1,1\}^B\ \mbox{where}\ \textbf{w}^{(ki)}=\textbf{w}_{i}-\textbf{w}_{k}\nonumber\\
& \mbox{and}\ {y}^{(ki)}={y}_{ki}+s_i-1\nonumber,\ \mbox{and}\ s_i\geq0\ \mbox{is a slack variable}.
\end{align}
Therefore, revising (8), we get
\begin{eqnarray}
   & \max_{\textbf{b}_i} \textbf{b}_i^{\intercal}\Big(F({\bf x}_i)+\frac{1}{2\mathcal P}\sum_{k=1}^c\textbf{w}^{(ki)}\Big) & \\
&\mbox{subject to} \ \textbf{b}_i \in\{-1,1\}^B, & \nonumber
\end{eqnarray}
which has the optimal solution at (Algorithm~1, Step 13) \[\textbf{b}_i={\rm sgn}\left(F({\bf x}_i)+\frac{1}{2\mathcal P}\sum_{k=1}^c\textbf{w}^{(ki)}\right).\]}

\subsection{Theoretical Analysis of \textbf{Algorithm~1}}

Prior to presenting our theoretical analysis, we revise~\eqref{eqn:main2} to guarantee invariant regularization parameters with respect to code length $B$, class size $k$ and  $n$ sample instances. Then, the resulting objective function becomes
\begin{align}
    \label{eqn:main11}
& \min_{\mathbb B, \mathbb W, \mathbb H} \Big\{ \frac{\mathcal P'}{nB} \norm{\mathbb B-F(\mathbb X)}^2+ \frac{1}{nB}\norm{\mathbb B-\mathbb Y\mathbb W}^2 \cr 
& \qquad +\frac{\alpha'}{kB}\norm{\mathbb W}^2\Big\}.
\end{align}
Observe that~\eqref{eqn:main11} becomes~\eqref{eqn:main2}, when $\mathcal P'$=$\mathcal P$ and $\alpha'$=$\alpha k/n$. Now, we compute the computational complexity of Algorithm~1 and then develop its convergence analysis.

\textit{Computational Complexity}.
It is worth noting that our hashing-based SC is a lightweight approach (compared to deep learning frameworks~\cite{weng2021semantic, 9252948, 9398576, 9450827}). With our training set consisting of $n$ instances, where each instance is of $d$ dimension feature, and $k$ is the type of categories in the training set. Given $B$ bits code, the time complexity for learning: i) $\mathbb B$ is O\big({$ndB+nkB^2$}\big); ii) $\mathbb W$ is O\big({$ncB+nB^2+nk^2+B^3$}\big); and iii)  $\mathbb Q$ is O\big({$ndB+2nd^2$}\big). Therefore, total training time complexity is O\big({$ndB+nkB^2+B^3$}\big).

Next, we have the following convergence analysis of Algorithm~1.
\begin{mytheorem}\label{theorem:1}
In the $t^{\rm th}$ iteration, 
for given ${\mathbb B}_t$, Algorithm 1 converges when learning ${\mathbb W}_t$. After the $t^{\rm th}$ iteration, let ${\mathbb W}_t({\mathbb X}_1)$ and ${\mathbb W}_t({\mathbb X}_2)$ are learned by using  ${\mathbb X}_1$ and ${\mathbb X}_2$ respectively, then assuming $\norm{\textbf{b}- \textbf{y}\mathbb W^\intercal}\leq \text{\c c}$, where $\text{\c c}$ is a constant, we have
\begin{equation}
   \norm{\mathbb W_t(\mathbb X_1)-\mathbb W_t(\mathbb X_2)}\leq\frac{2k{\c c}}{\alpha'n}.
\end{equation}
\end{mytheorem}
\begin{IEEEproof}
We apply the properties of Bergman divergence, denoted by 
$\mathcal D_g(\mathbb X, \mathbb Y)$, and its extension~\cite{dhillon2008matrix} to prove Theorem~\ref{theorem:1}. Given a differentiable convex function that maps matrices to the extended real numbers, the Bergman divergence is always additive and non-negative. Mathematically,
$\mathcal D_g(\mathbb X, \mathbb Y)$ can be defined as
\cite{dhillon2008matrix}
\begin{eqnarray}
   \mathcal D_g(\mathbb X, \mathbb Y):=g(\mathbb X)-g(\mathbb Y)-\langle\nabla_g(\mathbb X), \mathbb X-\mathbb Y\rangle ,
\end{eqnarray}
where the inner product $\langle\mathbb X,\mathbb Y\rangle
=\Re \{ {\rm Tr}(\mathbb X\mathbb Y^{\star})\}$.

In the $t^{\rm th}$ iteration, let
\begin{align*}
   f_{\mathbb X}(\mathbb W)&=\frac{\mathcal P'}{nB} \norm{\mathbb B_t-F_{t-1}(\mathbb X)}^2+ \frac{1}{nB}\norm{\mathbb B_t-\mathbb Y\mathbb W}^2\cr &\quad +\frac{\alpha'}{kB}\norm{\mathbb W}^2 \cr
   g_{\mathbb X}(\mathbb W)&=\frac{\alpha'}{kB}\norm{\mathbb W}^2.
\end{align*}
Applying Bergman divergence with $f_{\mathbb X}(\mathbb W)$, after simplification, it can be shown that
\begin{align}
   &\mathcal D_f{_{\mathbb X_1}}(\mathbb W_t(\mathbb X_2),\mathbb W_t(\mathbb X_1))+\mathcal D_{f_{\mathbb X_2}}(\mathbb W_t(\mathbb X_1),\mathbb W_t(\mathbb X_2))\cr
   &=\frac{1}{nB}\Big[\norm{\textbf{b}_i- \textbf{y}_i\mathbb W_t(\mathbb X_1)^\intercal}^2-\norm{\textbf{b}_i- \textbf{y'}_i\mathbb W_t(\mathbb X_1)^\intercal}^2\cr
   &\quad +\norm{\textbf{b}_i- \textbf{y'}_i\mathbb W_t(\mathbb X_2)^\intercal}^2-\norm{\textbf{b}_i- \textbf{y}_i\mathbb W_t(\mathbb X_2)^\intercal}^2\Big]\cr 
   &\leq\frac{2{\text{\c c}}}{nB}\norm{\textbf{y'}_i\mathbb W_t(\mathbb X_1)^\intercal-\mathbb W_t(\mathbb X_2)}\cr
   & \quad+\frac{2{\text{\c c}}}{nB}\norm{\textbf{y}_i\mathbb W_t(\mathbb X_1)^\intercal- \mathbb W_t(\mathbb X_2)}\cr
     &\leq\frac{4{\c{c}}}{nB} \norm{\mathbb W_t(\mathbb X_1)- \mathbb W_t(\mathbb X_2)}.
\end{align}

Similarly, by applying Bergman divergence with $g_{\mathbb X}(\mathbb W)$ after simplification
\begin{eqnarray}
   &&\mathcal D_{g_{\mathbb X_1}}(\mathbb W_t(\mathbb X_2),\mathbb W_t(\mathbb X_1))+\mathcal D_{g_{\mathbb X_2}}(\mathbb W_t(\mathbb X_1),\mathbb W_t(\mathbb X_2))\nonumber\\
   &&=\frac{2\alpha'}{kB}\norm{\mathbb W_t(\mathbb X_1)-\mathbb W_t(\mathbb X_2)}^2 .
\end{eqnarray}
And we know that~\cite{dhillon2008matrix},
\begin{eqnarray}
&&\mathcal D_{f_\mathbb {X_1}}(\mathbb W_t(\mathbb X_2),\mathbb W_t(\mathbb X_1))+\mathcal D_{f_{\mathbb X_2}}(\mathbb W_t(\mathbb X_1),\mathbb W_t(\mathbb X_2))\nonumber\\ && \geq\mathcal D_{g_{\mathbb X_1}}(\mathbb W_t(\mathbb X_2),\mathbb W_t(\mathbb X_1))+\mathcal D_{g_\mathbb {X_2}}(\mathbb W_t(\mathbb X_1),\mathbb W_t(\mathbb X_2)), \nonumber  
\end{eqnarray}
therefore,
\begin{eqnarray}
\frac{4{\text{\c c}}}{nB}\norm{\mathbb W_t(\mathbb X_1)-\mathbb W_t(\mathbb X_2)}\geq\frac{2\alpha'}{kB}\norm{\mathbb W_t(\mathbb X_1)-\mathbb W_t(\mathbb X_2)}^2,\nonumber
\end{eqnarray}
which proves that
\begin{equation}
   \norm{\mathbb W_t(\mathbb X_1)-\mathbb W_t(\mathbb X_2)}\leq\frac{2k{\text{\c c}}}{\alpha'n}.
\end{equation}
\end{IEEEproof}
It is clear that Theorem~\ref{theorem:1} states that $\norm{\mathbb W_t(\mathbb X_1)-\mathbb W_t(\mathbb X_2)}$ (and $\norm{\mathbb B_t(\mathbb X_1)-\mathbb B_t(\mathbb X_2)}$) decreases with $n$.
In addition,
we can see that in each iteration the proposed loss $\norm{\mathbb B-\mathbb Y\mathbb W}^2$ mandates each signature $\textbf{b}_i$ of $\textbf{x}_i$ is from $\mathbb W$ using $\textbf{y}_i$. Also our proposed signature learning algorithm is stable with respect to $\mathbb W$. Our design of signature learning mandates that the gap $\norm{\mathbb W_t(\mathbb X_1)-\mathbb W_t(\mathbb X_2)}$ and the difference $\norm{\mathbb B_t(\mathbb X_1)-\mathbb B_t(\mathbb X_2)}$ decreases when $n$ increases and therefore the convergence of Algorithm~1 is guaranteed.
{\color{black}
\subsection{With Domain Adaptive Hashing (DAH)}
\label{sec:DAH}
With relevant insights from~\cite{zhang2022deep, ramakrishnan2020deep}, we extend our hashing-based SC approach for domain adaptation that includes unsupervised domain adaptation between the sender's and receiver's knowledge base, semantic extraction at the sender, and unsupervised hashing for the receiver. We employ Multi-Kernel Maximum Mean Discrepancy (MKMMD)~\cite{ramakrishnan2020deep} to quantify the distribution difference between sender and receiver datasets in a reproducing-kernel Hilbert space which  help reducing knowledge base disparity through nonlinear data alignment. In particular, with the extended framework, we seek to minimize (slight abuse of notations to distinguish sender and receiver and extending the ideas from~\cite{ramakrishnan2020deep})
\begin{eqnarray}
\min_{\mathbb X\in\{\mathbb X_s\cup \mathbb X_r\}}\mathcal J=F(\mathbb X_s) +\eta H(\mathbb X_s,\mathbb X_r) +\gamma DA(\mathbb X_s,\mathbb X_r)
\end{eqnarray}
where ($\eta,\gamma$) pair decides the importance of the entropy loss, $ H(\mathbb X_s,\mathbb X_r)$ and the domain adaptation, $DA(\mathbb X_s,\mathbb X_r)$. We follow the heuristics explained in~\cite{ramakrishnan2020deep} to set ($\eta,\gamma$) in our evaluation. 

Note that that hash values in the extended approach are computed from the output of
the whole network, and $\mathbb H= {\rm sgn} (F(\{\textbf{x}_s\cup \textbf{x}_r\})$ is the extended Hash function for encoding $\textbf{x}$ by  $B$ bits. With $\mathbb X_s=\{\textbf{x}_s\}_1^{n_s}$ and $\mathbb X_r=\{\textbf{x}_r\}_1^{n_r}$, mathematically, $ H(\mathbb X_s,\mathbb X_r)$ and $DA(\mathbb X_s,\mathbb X_r)$ can be computed by using~\eqref{eqn:entr} and~\eqref{eq:proj} respectively
\begin{eqnarray}
H(\mathbb X_s,\mathbb X_r)=-\frac{1}{n_r}\sum_{i=1}^{n_r}\sum_{j=1}^{c}P(ij) \log_2 P(ij)\\
DA(\mathbb X_s,\mathbb X_r)=\sum\norm{\mathbb E[\phi(\mathbb X_s)]-\mathbb E[\phi(\mathbb X_r)]}^2_{\mathscr{H}}
\end{eqnarray}
where, $\norm{\mathbb E[\phi(\mathbb X_s)]-\mathbb E[\phi(\mathbb X_r)]}^2_{\mathscr{H}}$ is the MKMMD associated with the corresponding layer represented by the kernel Hilbert space $\mathscr{H}$ of the resultant kernel $\langle\phi(\mathbb X_s), \phi(\mathbb X_r)\rangle$.}

\subsection{Practical Benefits of the Proposed SC Framework}
Applications that rely on real-time decision-making, such as smart grids and networked control, need a major rethink in their communication processes. There is an increasing interest in the investigation of an efficacious SC system in 6G that can jointly optimize information collection, distribution, and decision-making policies for a large scale autonomous network systems~\cite{pokhrel2022learning, uysal2021semantic}. 

{\color{black}
While such network systems are not-so-futuristic, the theoretical foundation for SC is poor since ``the proper or meaningful piece of information" is neither specified in classical data communication, nor is the process of information generation considered in the communication process. For the last decade,
\textit{our objective had been to ultra-reliably deliver an entire data stream as quickly as practicable}~\cite{uysal2021semantic}. However, such a classic strategy can indeed be extremely inefficient for 6G systems. Let us analyze the following two beneficial cases for 6G
with \textit{intelligent transportation systems} and \textit{industry 4.0} systems.

\begin{itemize}
\item Case 1: Negotiating for timely consensus among autonomous cars regarding intended moves is critical for preventing accidents in \textit{intelligent transportation systems}. This is critical in unexpected situations, such as the abrupt presence of pedestrians. The proposed hashing based SC approach, considering the time-dependent value of messages help the underlying network prioritize information flow while fulfilling safety requirements. An efficient and low latency signature learning mechanism (developed in Algorithm~1) enables the transmission of signatures along with message, which can build the foundation towards the desired goal-oriented communication.

\item Case 2: In \textit{industry 4.0 network systems}, plant statuses are frequently relayed to the remote server for intelligent manufacturing and process estimation. Communication between plants and servers, on the other hand, might experience unforeseen delays, not desirable in 6G. In this scenario, it has already been demonstrated that employing process-aware sparse sampling rather than uniform sampling can enhance performance by orders of magnitude [2]. This observation justifies our rationale behind designing Algorithm~1 for learning signatures (which can perform context-aware sampling) in 6G and piggybacking them with messages for the desired joint optimization across three levels. Such an approach can dynamically maximize 6G networks' utility, significantly reduce the sampling rate and the corresponding energy consumption on the plant side, which are vital for low-power and energy harvesting sensors over the industry 4.0 systems. 
\end{itemize}
}

\section{Performance Evaluation}

We perform experiments and evaluate the proposed hashing-based SC in terms of computational complexity, efficiency, and reconstruction performance at the receiver end. We have first utilized CIFAR-10. CIFAR-10 consists of 60k plus images categorized into ten classes, where each class consists of six thousand image samples. In our setup, we segregate the datasets into two parts, where five thousand samples are used for testing and the remaining for the training,
which are large image data sets tested extensively in the literature. The samples taken from CIFAR-10 data are normalized based on our requirements in data size (e.g., $B$ in bits). We set $\alpha=1$, maximum number of iterations $\mathcal I=100$, and $\mathcal P=0.0001$.

Several metrics have been developed with a view to assessing the code generated by the hashing methods~\cite{xia2014supervised, jiang2019evaluation}. The precision and mean average precision (MAP) based on the (Hamming)  radius $r$ are the most celebrated ones~\cite{xia2014supervised}. Precision at radius $r$ calculates the accuracy of the returned points whose Hamming distance, $dist_H$, is less than $r$. MAP evaluates the rank by averaging the precision at each position in the ranked list produced based on Hamming ranking. Considering ground-truth neighbors set $\mathbb Y_j=\{\textbf{x}_i|\forall \textbf{x}_i, y_{ji}=1\}$ and a Hamming ball $$
\mathbb B^{(r)}_j=\{\textbf{x}_i\ |\ \forall \textbf{x}_i, dist_H=r\},
$$
we compute $Y_j^+=|\mathbb Y_j|$, $Y_{j,r}=|\mathbb B^{(r)}_j|$ and $ Y_{j,r}^+=|\mathbb Y_j \cap \mathbb B^{(r)}_j|$.   Mathematically, given $\mathcal N$ retrieval queries~\cite{xia2014supervised, jiang2019evaluation}
\begin{align}
{\rm precision}(r) &= \frac{1}{\mathcal N}  \sum_{j=1}^{\mathcal N}\Big({\sum_{r}{ Y_{j,r}^+}}\Big/{\sum_{r}{ Y_{j,r}}}\Big);\nonumber\\
{\rm MAP} &= \frac{1}{\mathcal N}  \sum_{j=1}^{\mathcal N}\frac{1}{ Y_j^+}\sum_{k=1}^n{{\rm precision}(k)}{\mathbbm 1}(y_{ki}=1).\nonumber
\end{align}

We have observed the results in precision (lookup of $dist_H$=2) and MAP. To this end, ground truths are determined from the data sets by using specific category information, and if there is no return while calculating precision, it is considered a false case.

\begin{figure}
\centering
\begin{subfigure}{0.34\textwidth}
\includegraphics[width=1\linewidth]{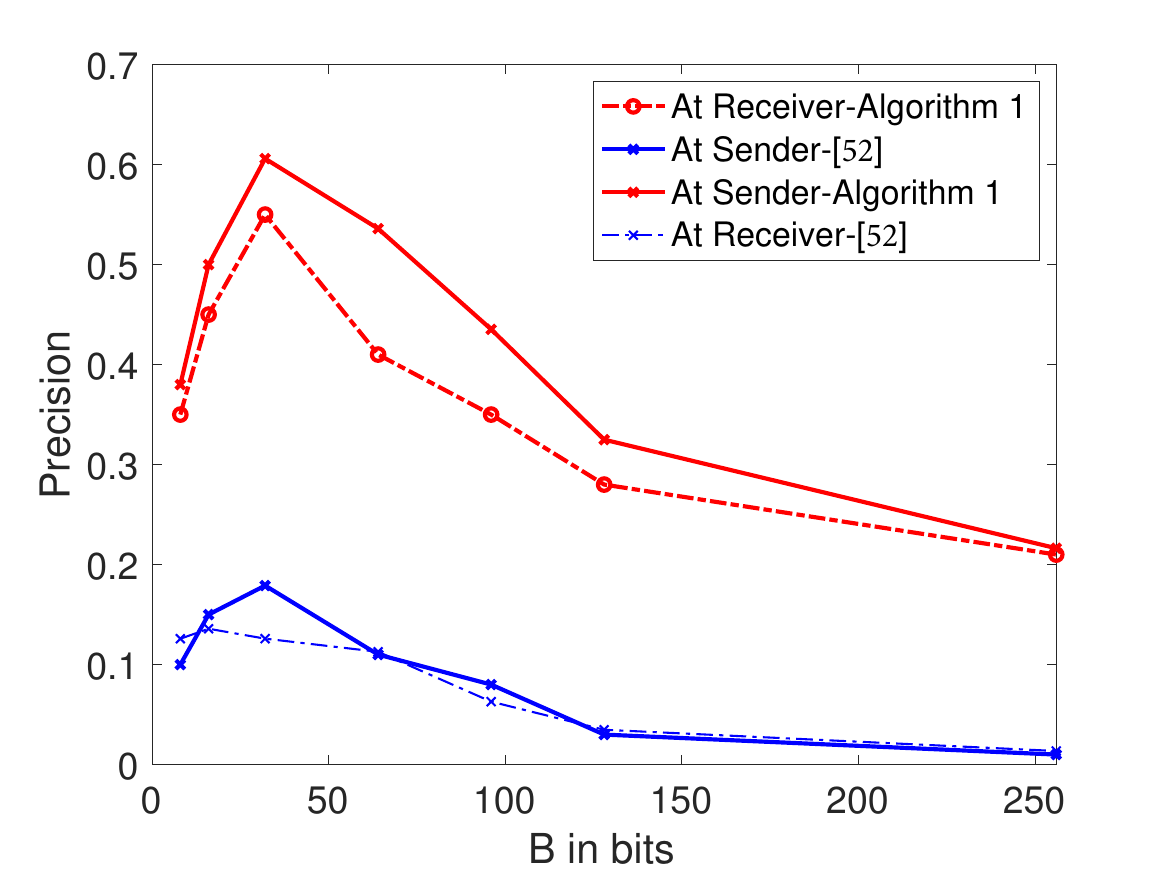}
\caption{precision of semantics reconstruction}
\vspace{5 mm}
\end{subfigure}
\begin{subfigure}{0.34\textwidth}
\includegraphics[width=1\linewidth]{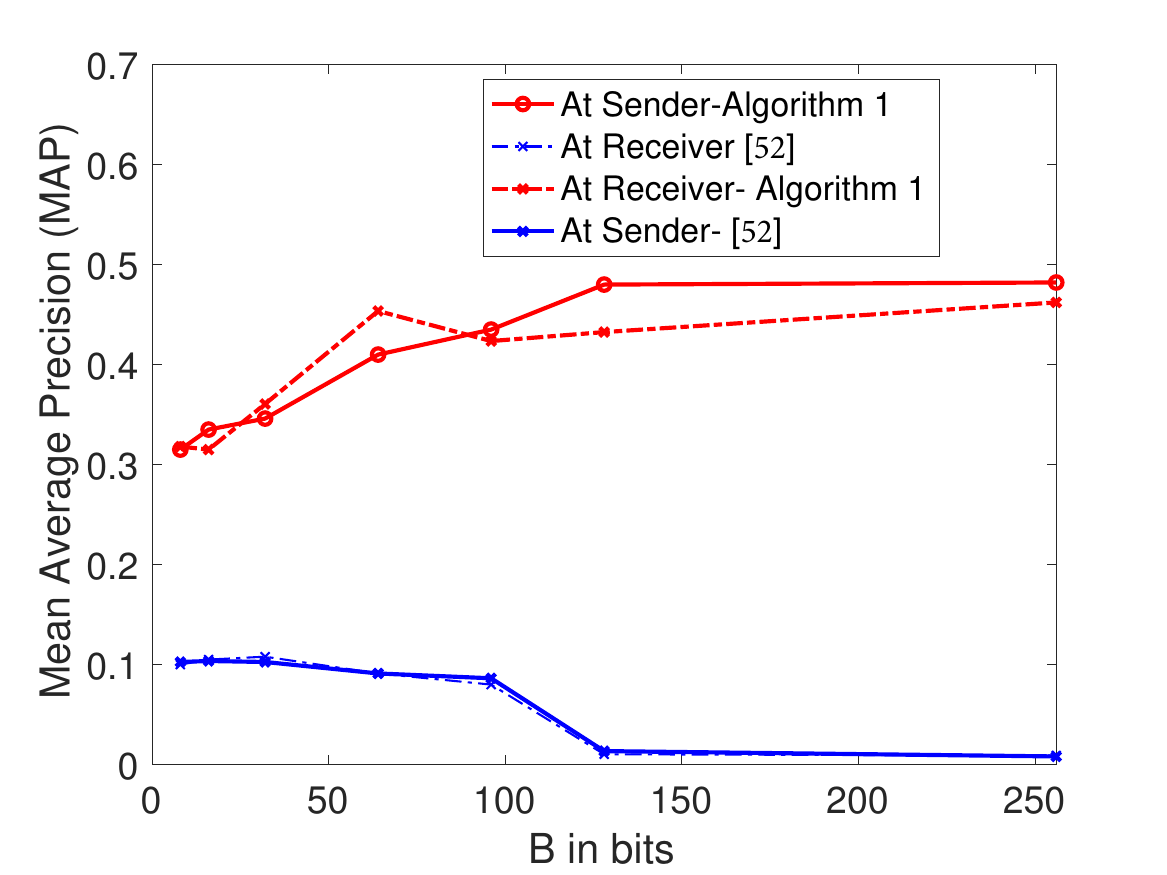}
\caption{MAP of semantics reconstruction}
\vspace{5 mm}
\end{subfigure}
\begin{subfigure}{0.33\textwidth}
\includegraphics[width=1\linewidth]{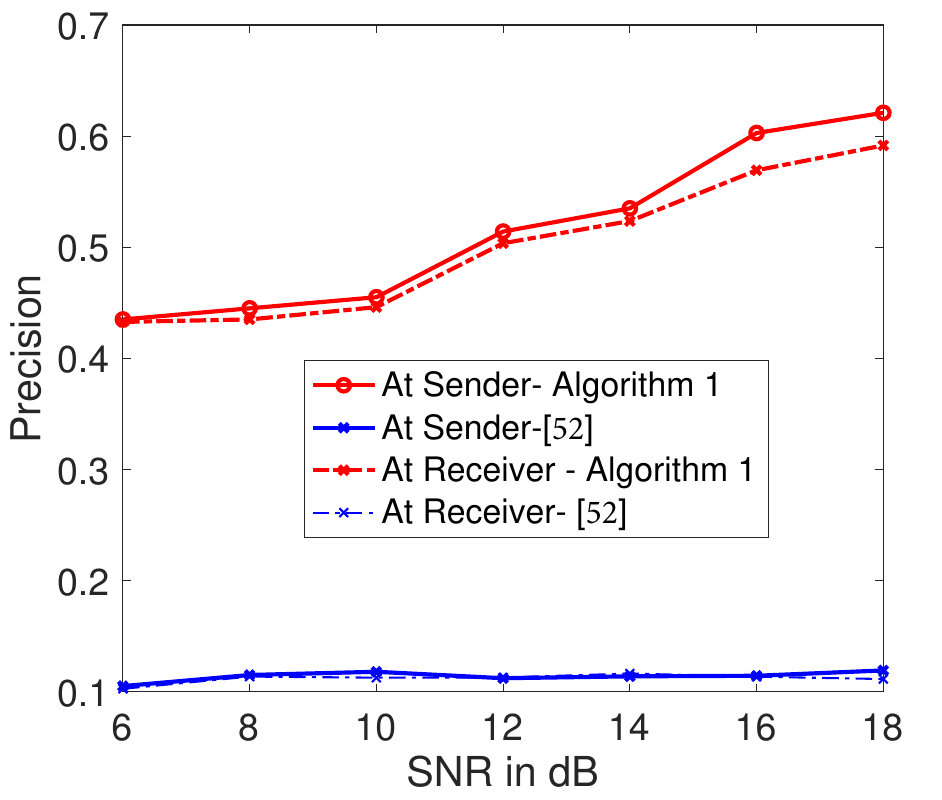}
\caption{precision of semantics reconstruction}
\vspace{5 mm}
\end{subfigure}
\begin{subfigure}{0.34\textwidth}
\includegraphics[width=1\linewidth]{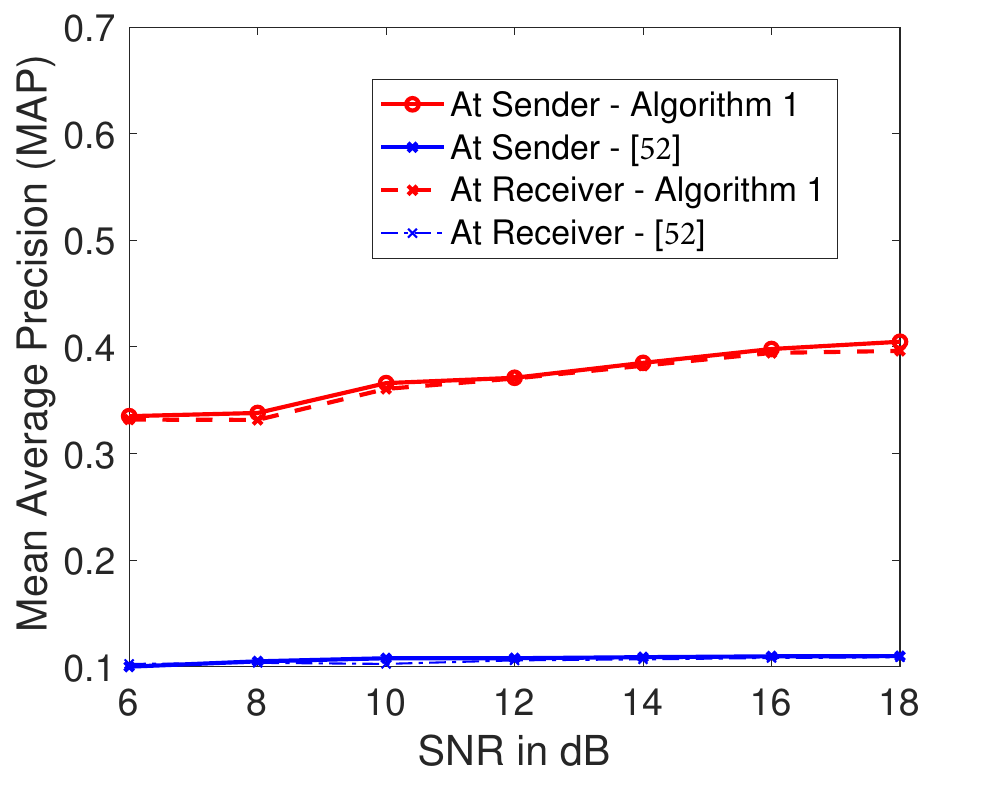}
\caption{MAP of semantics reconstruction}
\vspace{7 mm}
\end{subfigure}
\caption{Precision and MAP ($dist_H$ is 2) with the models trained under
AWGN channel and compared in terms of 
semantics signature reconstruction at the receiver. In a) and b), SNR is fixed at 10dB, $B$ ranges from 16 bits to 256 bits. However, in c) and d), SNR varies from 6 dB to 18 dB, $B$ is fixed to 96 bits.\label{fig:fig1}}
\end{figure}

Our first channel model is an emulator that adds white Gaussian noise to the signals passing through it. This AWGN model is applied to each element of the whole input signal where the scalar value of SNR is introduced. We also consider Rayleigh and Rician channel for further evaluations. For a given SNR, the precision and MAP are compared with different lengths in $B$ (in bits), as illustrated in Figs.~\ref{fig:fig1} a) and~\ref{fig:fig1} b).



{\color{black}Figs.~\ref{fig:fig1} a) and~\ref{fig:fig1} b) show that the proposed algorithm achieves the closest match of precision and MAP at the receiver with ground truth at the sender. The purpose of evaluating both the sender and receiver sides is to quantify the impact of semantic noise and compare/observe the reconstruction accuracy. In addition, we have compared the results at both the receiver and sender with one of the well-known hashing methods~\cite{kulis2009learning} (the code is publicly available). The later comparison reveals the fact that for higher precision $\textbf{b}_i$ should be taken into account (recall from Sec. II-A, that solving~\eqref{eqn:main} is NP-hard with discrete $\textbf{b}_i$).} 

Figs.~\ref{fig:fig1} a) and~\ref{fig:fig1} b) provide an essential insight that one of the main benefits of the proposed hashing-based SC approach is its abstraction efficiency in encoding semantcis signatures. Perhaps one of the immediate advancements of the proposed approach could be designed towards generating a single semantic signature to represent even a video, a picture or a sound with much the same semantic, which seems to be invaluable for communication even with ultra-low bandwidth.


\begin{figure}
\centering
\begin{subfigure}{0.34\textwidth}
\includegraphics[width=1\linewidth]{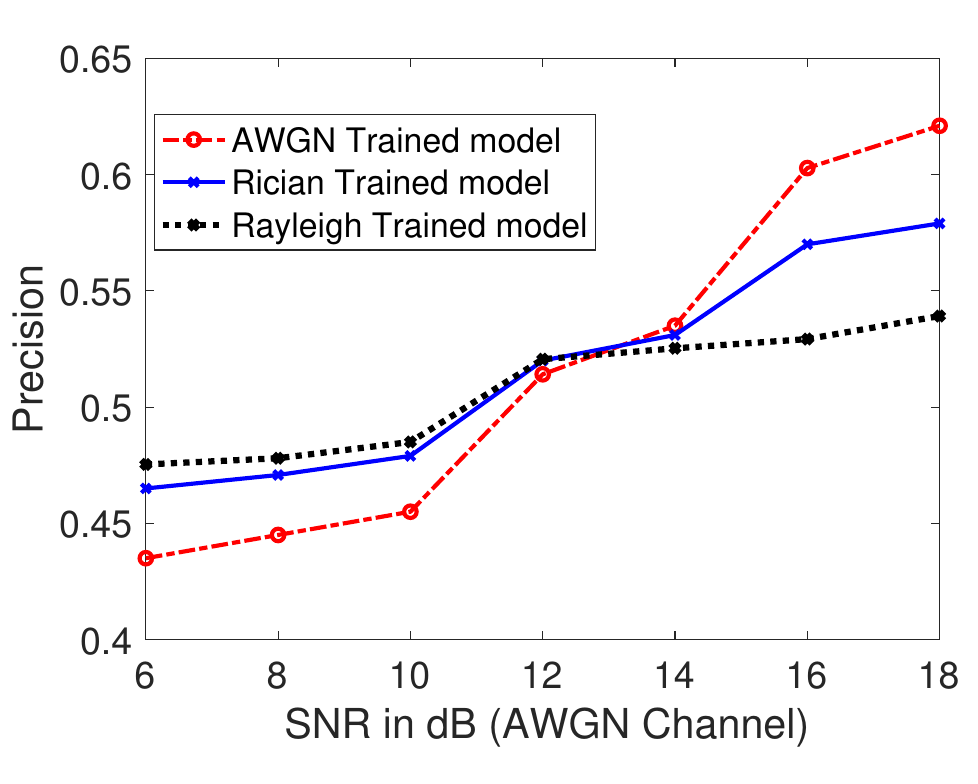}
\caption{Precision over AWGN Channel.}
\vspace{5 mm}
\end{subfigure}
\begin{subfigure}{0.33\textwidth}
\includegraphics[width=1\linewidth]{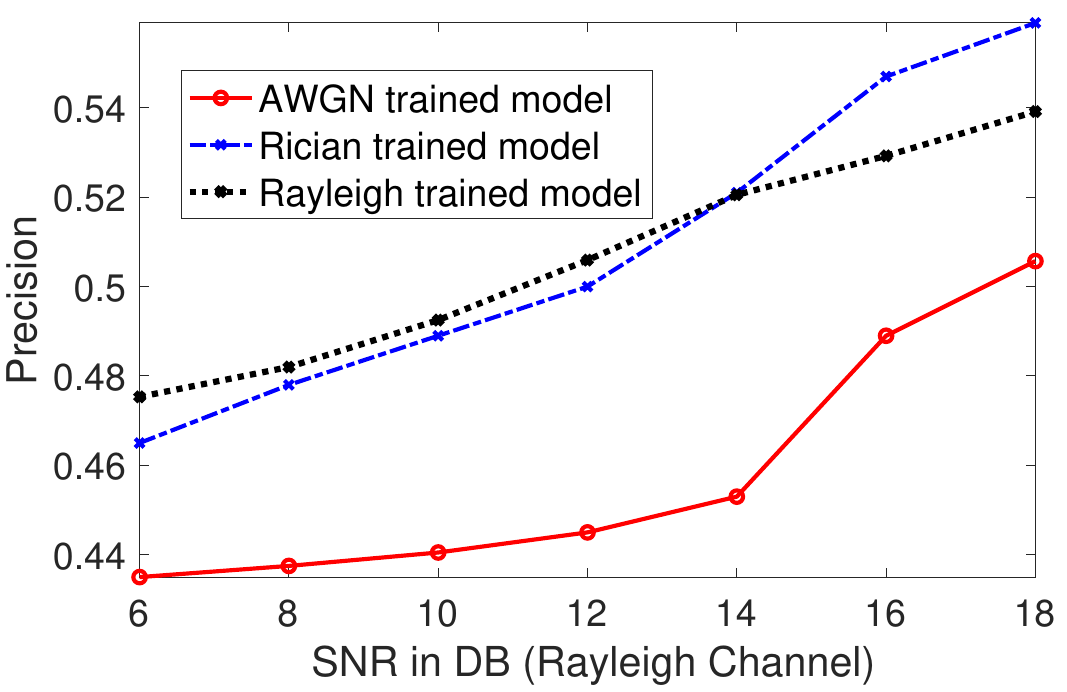}
\caption{Precision over Rician Channel.}
\vspace{5 mm}
\end{subfigure}
\begin{subfigure}{0.33\textwidth}
\includegraphics[width=1\linewidth]{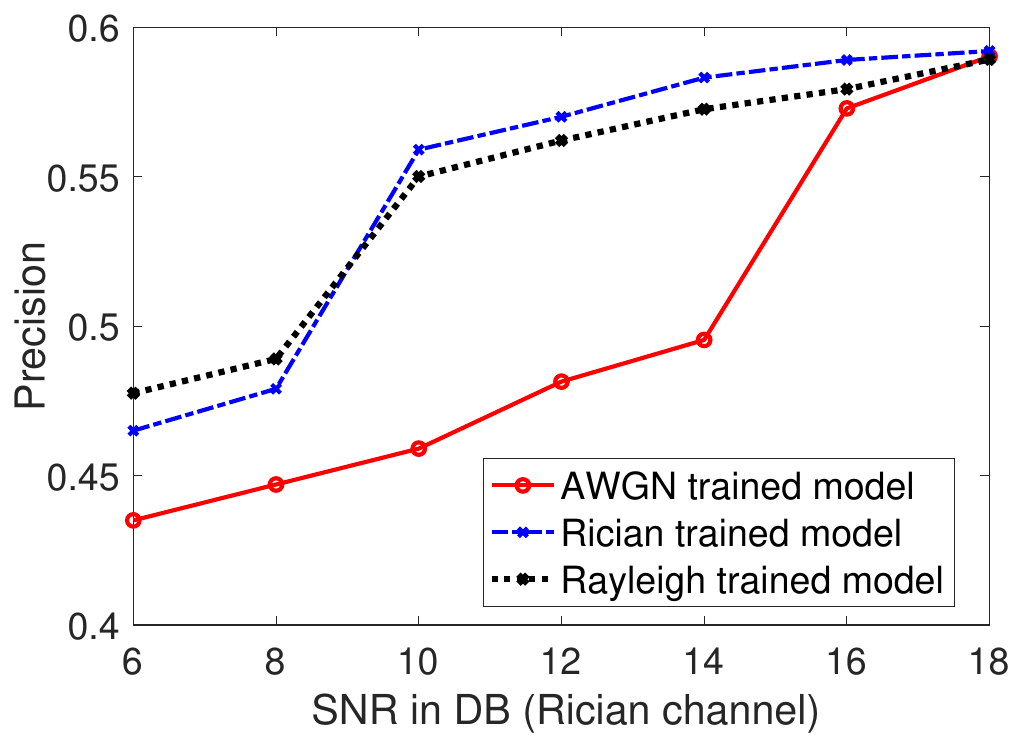}
\caption{Precision over Rayleigh Channel.}
\vspace{5 mm}
\end{subfigure}
\begin{subfigure}{0.32\textwidth}
\includegraphics[width=1\linewidth]{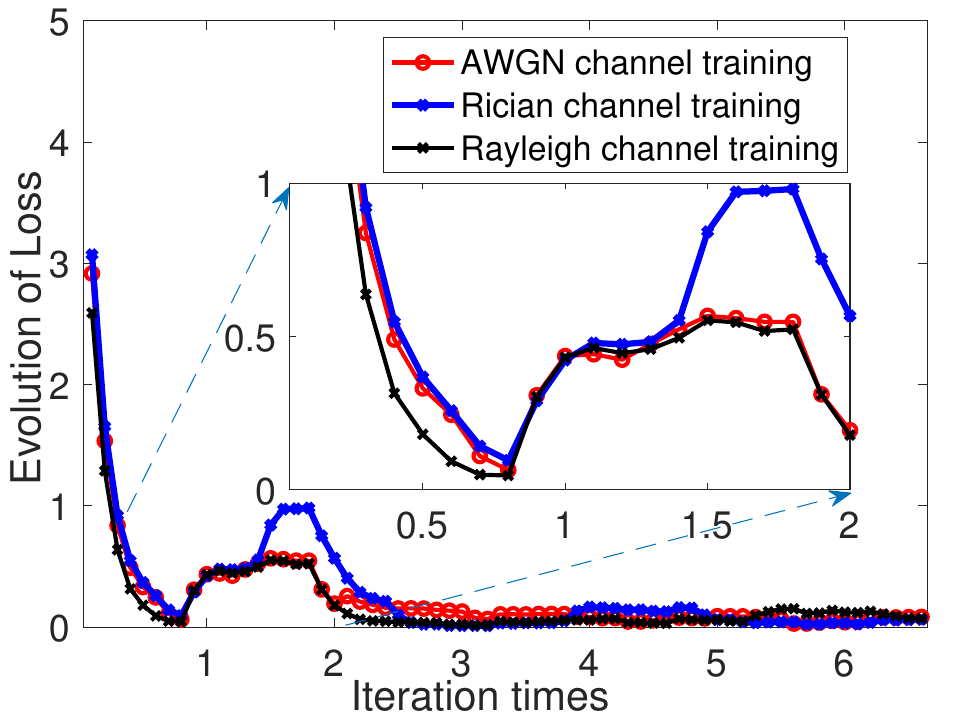}
\caption{Convergence dynamics of Algorithm 1 (loss vs iteration times).}
\vspace{5 mm}
\end{subfigure}
\caption{Precision and convergence of Algorithm 1 ($B$ is 96 bits and $dist_H$ is 2) with the models trained under
various channels and evaluated under AWGN, Rician and Rayleigh channel and the Convergence of Algorithm 1 (loss evolution vs iteration times).\label{fig:fig0}}
\end{figure}
\begin{figure}[t]
\centering
\begin{subfigure}{0.34\textwidth}
\includegraphics[width=1\linewidth]{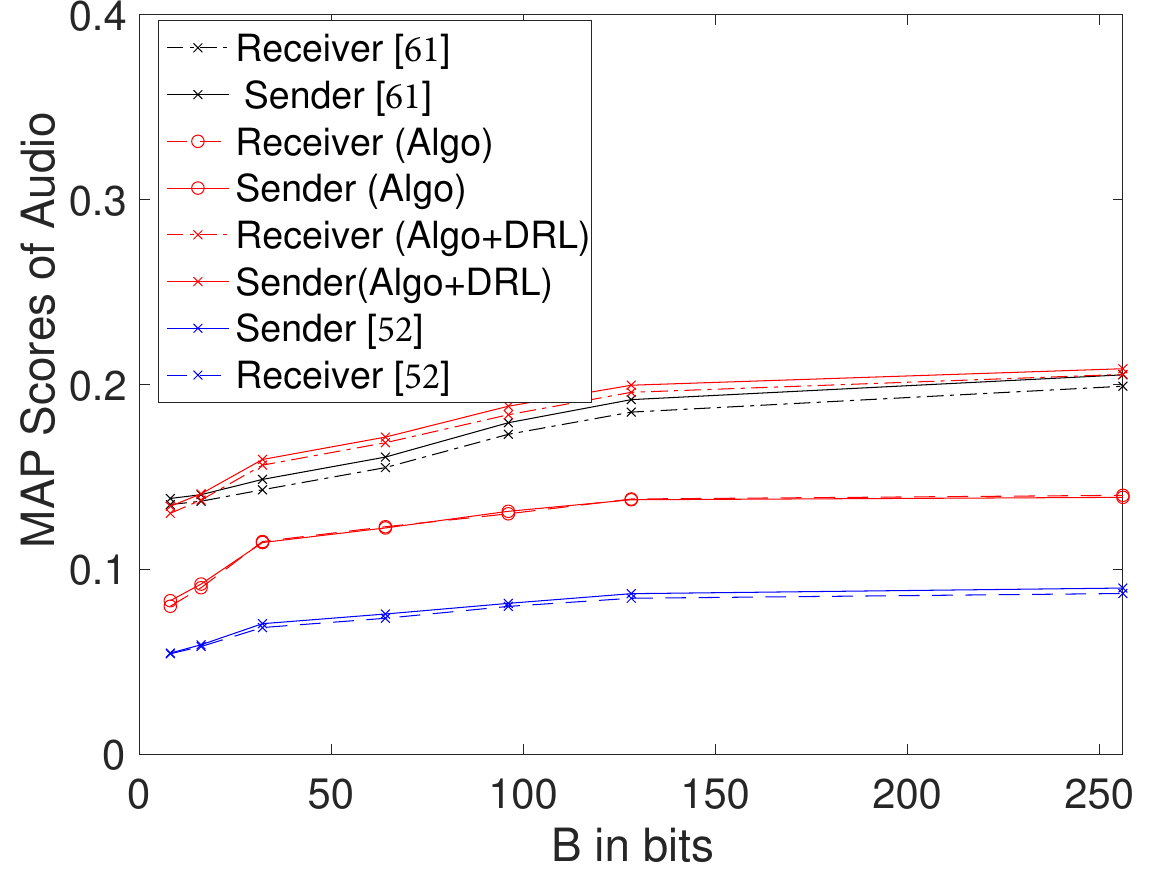}
\caption{MAP scores of the audio compared with the  semantics reconstruction}
\vspace{5 mm}
\end{subfigure}
\begin{subfigure}{0.34\textwidth}
\includegraphics[width=1\linewidth]{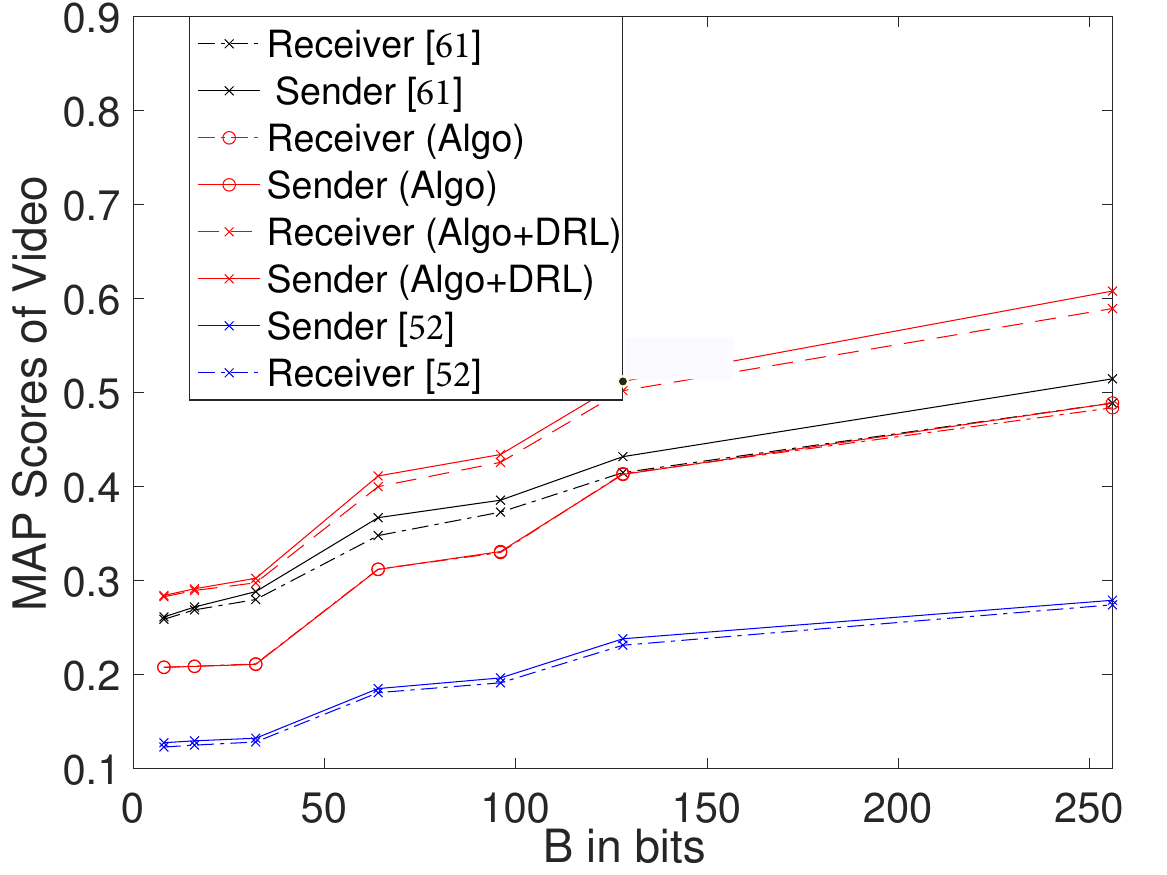}
\caption{MAP scores of the video compared with the semantics reconstruction}
\vspace{5 mm}
\end{subfigure}
\caption{\color{black} MAP ($dist_H$ is 2) with the models trained under
AWGN channel and compared in terms of 
semantics signature reconstruction at the receiver. In a) and b), SNR is fixed at 10dB, $B$ ranges from 16 bits to 256 bits. \label{fig:fig4}}
\end{figure}
Furthermore, by fixing $B= 96$ bits, 
the precision and MAP are evaluated with varying SNR of the channel (using AWGN channel model), as illustrated in Figs.~\ref{fig:fig1} c) and~\ref{fig:fig1} d). {\color{black}As noted in Fig.~\ref{Fig:system}, we found that the proposed algorithm achieves the closest precision match in reconstructing semantics and the MAP at the receiver end compared with the ground truth under similar channel conditions. Such semantic reconstruction has an intrinsic capability to offer multimodal experiences in 6G networks, facilitating the transfer of skills over the Internet. We anticipate that such a framework may eventually lead to the emergence of the Internet of Senses (IoS), combining visual, auditory, haptic, and other technologies that enable people to experience feelings/senses remotely.}

Extensive evaluations with the models trained under
various physical channels (AWGN, Rician, Rayleigh) are shown in Fig.~\ref{fig:fig0}. Observe in Fig.~\ref{fig:fig0} a) that the precision of the three models
is evaluated over AWGN. The model trained with AWGN outperforms others when SNR
is high ($>14$ dB). In Fig.~\ref{fig:fig0} b) model trained with AWGN appears
poor over the Rayleigh channel. Furthermore, Fig.~\ref{fig:fig0} c) demonstrates their comparison over the Rician channel. Overall, we found that the model trained with the Rician channel setup is resilient to
different environments. Fig.~\ref{fig:fig0} d) demonstrates the convergence dynamics of our Algorithm 1 under other channel conditions. The dampings are expected because Algorithm 1 strictly updates each bit (iteratively) till convergence with a better set of semantics signatures, i.e., a set of binary codes $\mathbb B=\{\textbf{b}_i\}_1^n$. In our evaluations and experiments,
we noticed that the whole $B$ bits could be learned iteratively in $\tau B$ times, which usually converges when iteration times $\tau$ = 3 to 6.
{\color{black}\subsection{With Multimodal Dataset}
To further evaluate the proposed framework, we used the publicly available multimodal dataset, PKU XMediaNet~\cite{peng2018modality}, comprising of five different media (text, image, audio, video and 3D models). To verify the hierarchical semantic structure, Peng \textit{et al.}~\cite{peng2018modality} applied 200 category nodes from WordNet and generated this dataset consisting of 48 kinds of animals and 152 types of artifacts. 

{\color{black}{Figs.~\ref{fig:fig4} a) and~\ref{fig:fig4} b) show that the proposed algorithm achieves the closest match of the MAP scores for the audio and the video at the receiver when compared with ground truth at the sender. In addition, we have compared the results at both the receiver and sender by generating signatures with~\cite{kulis2009learning} and one of the well-known generative adversarial (GAN) hashing methods~\cite{zhang2019multi}. Only for the purpose of benchmarking and to provide a fair assessment, we apply deep reinforcement learning (DRL) features with hashing methods in our experiments. In our preliminary analysis, we have observed that when our algorithm is integrated with DRL (along the lines of that of~\cite{peng2019deep}), it outperforms~\cite{zhang2019multi} in both cases.  In our numerical experiments, we found that ``Algo+DRL" improved robustness in the low
SNR regime, however, due to the lack of differentiable loss
function mentioned earlier and real-time communication requirements in the anticipated SC for 6G systems, the adoption of deep hashing-based signatures remains questionable.}}

\subsection{Towards Sophisticated Hashing-based Signatures}

Our primary focus in this work is to quantify the category information by using semantics signature for the regression or classification to attain the proposed SC by design and therefore we assume negligible correlation between data samples. With relevant insights from this work and by adopting enhanced hashing approaches~\cite{li2017deep, zhang2021probability, peng2019deep, zhang2019multi} (studied in a different contexts) an emerging important research direction is towards developing improved signature by capturing the pairwise correlations between samples or by  jointly considering the correlations along with semantic reconstruction. For example, one may apply deep image hashing with a deep neural network to convert an input image to compact signatures, allowing efficient image retrieval over large-scale knowledge-base with huge datasets.  Deep hashing~\cite{li2017deep} has attracted increasing attention from the academic community as a result of the enormous increase in the current knowledge base and the data sets. However, the SC needs of proliferating real-time systems in different domains, is not supportable by these paradigms. To analyze this, we performed  retrieval efficiency comparison of the different existing methods in the literature. Convolutional neural networks (CNNs), such as ResNet, sit at the heart of deep hashing framework, however, there are several limitations for straightforward applications to SC. To provide a fair assessment, we compare deep learning features retrieved by the CNN network with typical hashing methods in our experiments. The histograms of the average computation time of different methods are shown in Figs.~\ref{fig:fig6} a) and~\ref{fig:fig6} b). Peng \textit{et~al.}~\cite{peng2019deep} explained insights toward improving hash function correlation by combining deep reinforcement learning with a sequential learning technique.

\begin{figure}[t]
    \centering
    \includegraphics[scale=0.4]{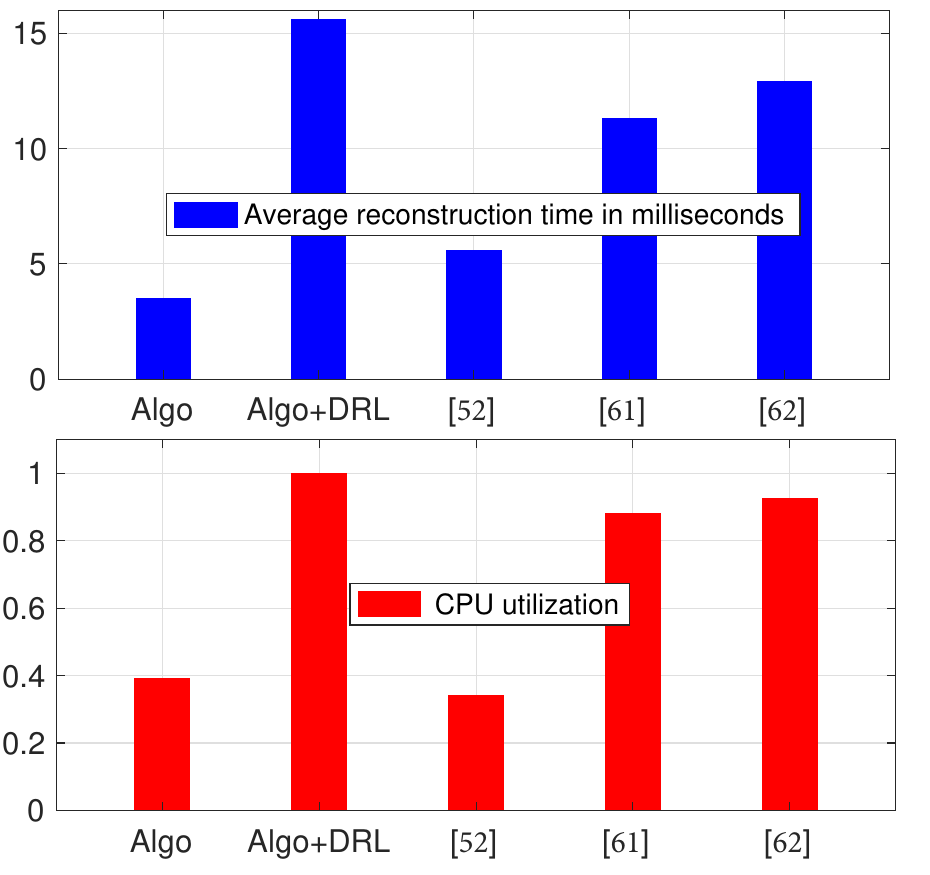}
    \caption{Comparison of CPU utilization in signature learning and the mean reconstruction time in milliseconds using CIFAR-10 by fixing the signature length to 64 bits.}
    \label{fig:fig6}
\end{figure}

Whilst generation of the results in Fig.~\ref{fig:fig6} we have found that CNNs perform well when categorizing very similar images to the dataset. However, if the images contain some degree of tilt or rotation, CNNs typically struggle to classify the image. Such observation has already been justified in the literature~\cite{sabour2017dynamic}. With CNNs, neurons provide all minute detail recognition in the convolutional layer. These high-level neurons then determine if all traits are there. The image is strided to determine whether or not features are present. During this phase, the CNN may lose the knowledge of the composition and position of the components in the image and sends the information to a neuron, which often cannot categorize them effectively.  

Fig.~\ref{fig:fig6} b) shows the comparison of CPU utilization in signature learning and the mean reconstruction time in milliseconds using CIFAR-10 by fixing the signature length to 64 bits. When comparing with other methods~\cite{kulis2009learning, zhang2019multi, peng2019deep}, in Fig.~\ref{fig:fig6}, our algorithm outperforms with approximately 40\% CPU utilization while learning signatures and the average retrieval time is below 5~ms. Due to aforementioned limitations, and the results in Fig.~\ref{fig:fig6},  we find that existing deep hashing are not directly applicable for the SC paradigm (freshness, age-of-information and the extreme low latency requirements of 6G~\cite{uysal2021semantic}). 
\begin{figure}[t]
\centering
\begin{subfigure}{0.34\textwidth}
\includegraphics[width=1\linewidth]{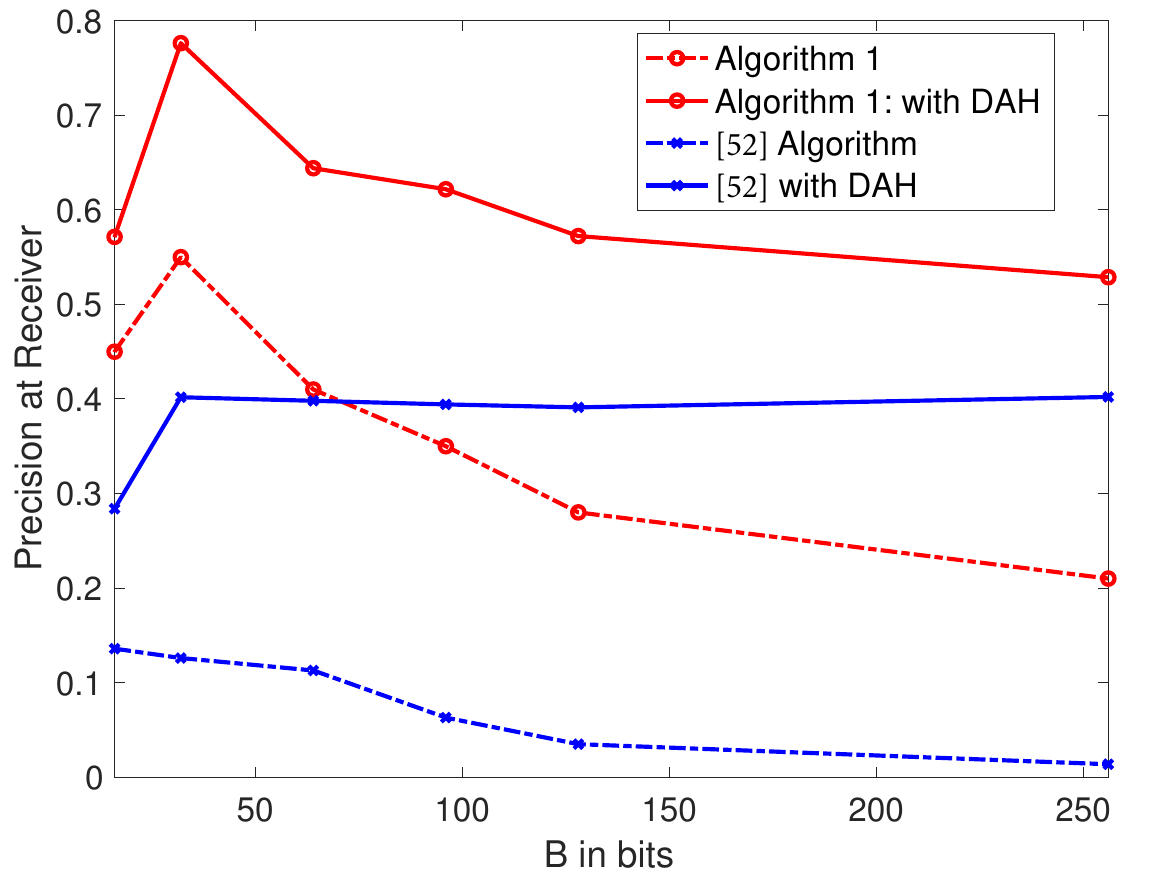}
\caption{Precision of the  semantics reconstruction at the receiver with and without DAH (recall Fig.~\ref{fig:fig1} (a)).}
\vspace{5 mm}
\end{subfigure}
\begin{subfigure}{0.34\textwidth}
\includegraphics[width=1\linewidth]{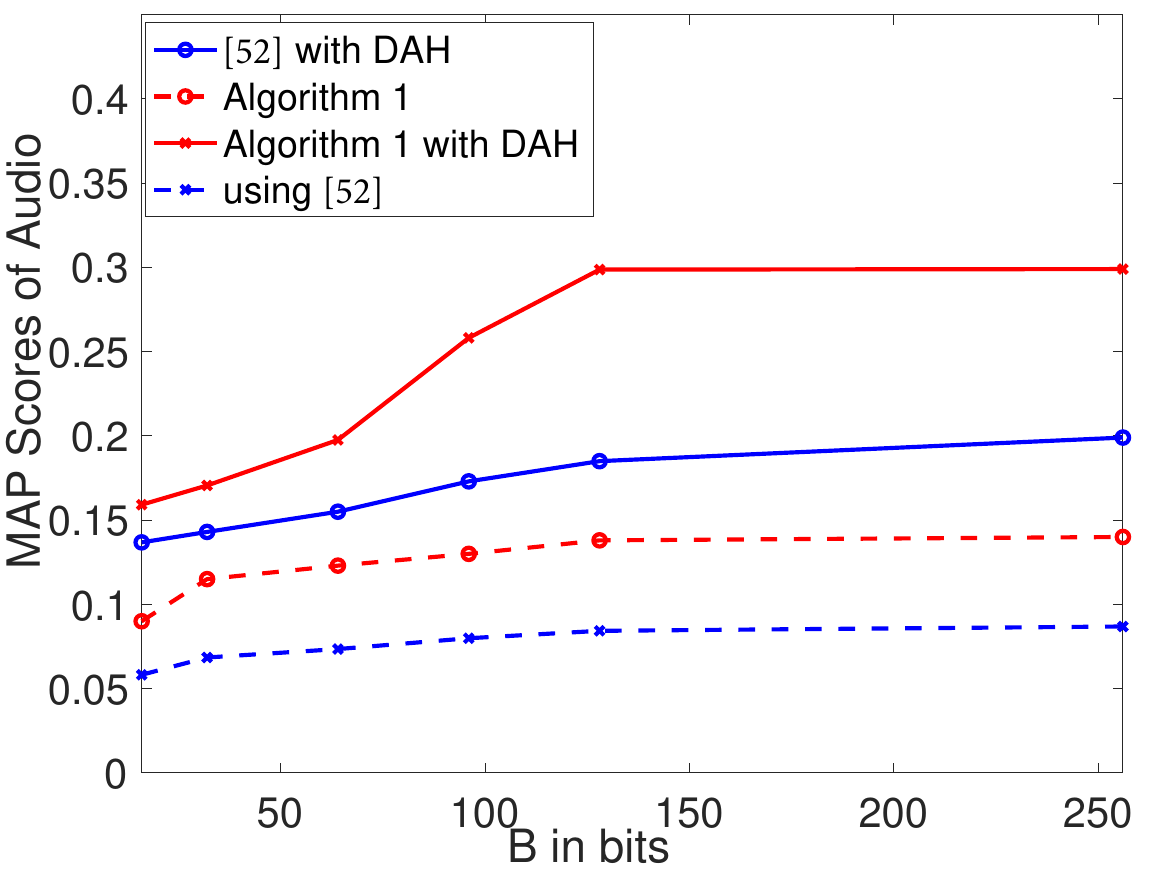}
\caption{Comparison of the MAP scores of the Audio with and without DAH (recall Fig.~\ref{fig:fig4} (a)).}
\vspace{5 mm}
\end{subfigure}
\caption{\color{black} Importance of Domain Adaptation and Entropy Loss: Comparing Fig. 3 (a) and Fig. 5 (a)
with and without DAH at the receiver end. \label{fig:figDAH}}
\end{figure}
Using the extended approach developed in Sec.~\ref{sec:DAH} (Algorithm ~1 with DAH), we revisit all important results under the same setting and network conditions.  In Fig.~\ref{fig:figDAH}, we show the results of Fig.~\ref{fig:fig1} (a) and Fig.~\ref{fig:fig4} (a) using DAH for comparison. Fig.~\ref{fig:figDAH} demonstrates that both the precision and MAP of semantic reconstruction at the receiver improves significantly with the proposed DAH. As anticipated in Sec.~\ref{sec:DAH}, these observations reflect the importance of the entropy loss, $ H(\mathbb X_s,\mathbb X_r)$ and the domain adaptation, $DA(\mathbb X_s,\mathbb X_r)$ in the accuracy of semantic extraction for the SC framework.

\section{Concluding Remarks}
 Semantic communication framework design is the major departure that is rapidly evolving to facilitate the real momentum towards 6G networks. It aims to jointly optimize \textit{information collection, distribution, and decision-making policies} in 6G systems. We proposed and evaluated a hashing-based SC framework, where the learning objective is to generate the optimal binary hash codes as semantics signatures for transforming semantics understanding along with the semantic message.  We have evaluated the extended framework with domain adaptation using large data sets and demonstrated its precision and effectiveness in image and multimedia transmission. In the future, further advancement of the proposed framework could harness the substantial advances in sensor and actuator technology, such as smell actuation and high-quality haptics.} 


\bibliographystyle{IEEEtran.bst}
\bibliography{References}
\end{document}